\documentclass[12pt,a4paper]{article}
\nonstopmode
\usepackage{enumitem}   
     
\usepackage{epsfig}
\usepackage[a4paper]{geometry}
\usepackage{float}
\usepackage[stable]{footmisc}
\usepackage[utf8]{inputenc}
\usepackage{a4wide}
\usepackage{amsmath,amssymb,amstext,amsthm,amssymb}
\usepackage{amsfonts,mathrsfs}
\usepackage{framed}
\usepackage{graphicx}
\usepackage{color}
\usepackage[most]{tcolorbox}
\usepackage{bbold}
\usepackage{bbm}
\usepackage[normalem]{ulem}
\usepackage{rotating} 
\usepackage{slashed} 
\usepackage{cancel} 
\usepackage{braket} 
\usepackage{amstext}
\usepackage{array}
\usepackage{setspace}
\usepackage[urlcolor=blue,colorlinks=true,linkcolor  = black]{hyperref}
\usepackage{overpic}
\usepackage{bbm}
\usepackage{cite}
\usepackage{ulem}
\usepackage{lipsum}
\usepackage{tikz}
\usepackage{mathtools}
\usepackage{amsthm}
\usepackage{mathrsfs}

\usepackage{fontawesome5}
\makeatletter
\newcommand{\github}[1]{%
   \href{#1}{{\normalsize \color{black}\faGithub}}%
}
\makeatother

\newcommand{\DeclareRuneSeparators}[1]{} 

\input{arm.fd}

\usepackage{tikz-cd}
\usepackage{tikz}
\usepgflibrary{arrows} 
\usetikzlibrary{arrows,shapes,positioning} 
\usetikzlibrary{decorations.markings}
\usetikzlibrary{shapes.geometric}
\usetikzlibrary{arrows.meta,arrows}

\tikzstyle arrowstyle=[scale=1]
\tikzstyle directed=[postaction={decorate,decoration={markings,
    mark=at position .5 with {\pgftransformscale{2}\arrow[arrowstyle]{stealth}}}}]

\let\baraccent=\=
\renewcommand{\=}[1]{\stackrel{#1}{=}}

\newcommand{\cT}{\mathcal{T}}

\newcommand{\kom}{\, ,\quad}

\DeclareSymbolFontAlphabet{\mathbb}{AMSb}

\renewcommand{\L}{\Lambda}

\def\bbZ{\mathbb{Z}}

\def\cC{\mathcal{C}}

\newcommand{\ignore}[1]{}

\usepackage{todonotes}

\begin{document}
	
	\pagestyle{plain}

	\makeatletter
	\@addtoreset{equation}{section}
	\makeatother
	\renewcommand{\theequation}{\thesection.\arabic{equation}}
	\pagestyle{empty}
	\begin{flushright}
        \text{LMU-ASC 06/24}
        \end{flushright}
	\vspace{0.5cm}
	 
	\begin{center}
		
		{{\LARGE \bf The DNA of Calabi-Yau Hypersurfaces}\\[7.5mm]
            {\large \emph{A Genetic Algorithm for Polytope Triangulations}}\\[12.5mm]
        }
	\end{center}

	\begin{center}
		\scalebox{0.95}[0.95]{{\fontsize{14}{30}\selectfont Nate MacFadden,$^{a}$ Andreas Schachner,$^{a,b}$ and Elijah Sheridan$^{a}$}}
	\end{center}

	\begin{center}
		\vspace{0.25 cm}
		\textsl{$^{a}$Department of Physics, Cornell University, Ithaca, NY 14853 USA}\\
        \textsl{$^{b}$ASC for Theoretical Physics, LMU Munich, 80333 Munich, Germany}\\
  
		\vspace{1cm}
		\normalsize{\bf Abstract} \\[8mm]
		
	\end{center}

	\begin{center}
		\begin{minipage}[h]{15.0cm}

            We implement Genetic Algorithms for triangulations of four-dimensional reflexive polytopes which induce Calabi-Yau threefold hypersurfaces via Batyrev's construction. We demonstrate that such algorithms efficiently optimize physical observables such as axion decay constants or axion-photon couplings in string theory compactifications. For our implementation, we choose a parameterization of triangulations that yields homotopy inequivalent Calabi-Yau threefolds by extending fine, regular triangulations of two-faces, thereby eliminating exponentially large redundancy factors in the map from polytope triangulations to Calabi-Yau hypersurfaces. In particular, we discuss how this encoding renders the entire Kreuzer-Skarke list amenable to a variety of optimization strategies, including but not limited to Genetic Algorithms. To achieve optimal performance, we tune the hyperparameters of our Genetic Algorithm using Bayesian optimization. We find that our implementation vastly outperforms other sampling and optimization strategies like Markov Chain Monte Carlo or Simulated Annealing. Finally, we showcase that our Genetic Algorithm efficiently performs optimization even for the maximal polytope with Hodge numbers $h^{1,1} = 491$, where we use it to maximize axion-photon couplings. Our methods for sampling and optimization are implemented in a Python package \texttt{cyopt}. \github{https://github.com/sheride/cyopt}

		\end{minipage}
	\end{center}
    \vfill
    \today
    \newpage
	
    \setcounter{page}{1}
    \pagestyle{plain}
    \renewcommand{\thefootnote}{\arabic{footnote}}
    \setcounter{footnote}{0}
    %
    %
    \setcounter{tocdepth}{3}
	\tableofcontents

\section{Introduction}
\label{Intro}

The string landscape of Effective Field Theories (EFTs) obtained from dimensional reductions of string theory is the perfect arena to study general properties of quantum gravity. Its computational complexity is only exacerbated by its vastness with estimates ranging up to $10^{272,000}$ \cite{Taylor:2015xtz}. Even though such numbers are expected to be vast overcountings, it is still believed that the actual number of viable EFTs in the string landscape is far too big to allow for a systematic enumeration.

A large fraction of this landscape can be attributed to the choice of compact geometries. Prominent examples are Calabi-Yau threefold hypersurfaces obtained from fine, regular, star triangulations (FRSTs) of four-dimensional reflexive polytopes as enumerated in the Kreuzer-Skarke (KS) database \cite{Kreuzer:2000xy}.\footnote{The relevant measure of the size of this `toric landscape' is the count of topological equivalence classes, as was recently studied for small $h^{1,1}$ in \cite{Chandra:2023afu,Gendler:2023ujl}. Currently and foreseeably, however, such methods appear limited to small $h^{1,1}$.} These can be efficiently constructed via toric methods \cite{Batyrev:1993oya} using software packages like \texttt{CYTools} \cite{Demirtas:2022hqf}. Compactifications of Type II superstring theory on such manifolds lead to $\mathcal{N}=2$ supergravity in four dimensions which is a common starting point for studying phenomenology in string theory, see \cite{Cicoli:2023opf,McAllister:2023vgy,Douglas:2023yof} for recent reviews.

In the absence of systematic guiding principles, explicit string theory constructions like KKLT \cite{Kachru:2003aw} are typically obtained via brute-force methods by choosing Calabi-Yau threefolds mostly at random, generating vast amounts of EFTs only to keep a small fraction adhering to the often stringent constraints of the construction. Clearly, this constitutes a rather inefficient use of computational resources which motivates exploring more targeted approaches via algorithmic strategies.

In this work, we use \emph{Genetic Algorithms} (GAs) to address such optimization problems over the space of FRSTs of a given 4D reflexive polytope $\Delta^\circ$. Inspired by natural selection processes, GAs explore pseudo-continuous fitness landscapes for optimal solutions to problems with vast input spaces. Since such energy landscapes are ubiquitous in science, GAs have been shown to be extremely successful in many areas in physics ranging from particle phenomenology \cite{Allanach:2004my,Akrami:2009hp,Abel:2018ekz} over astrophysics \cite{Metcalfe:2000xn,Mokiem:2005qf,Rajpaul:2012wu} to cosmology \cite{Nesseris:2012tt,Hogan:2014qsa,Abel:2022nje}. They are also powerful tools to understand the string landscape \cite{Blaback:2013ht,Damian:2013dq,Damian:2013dwa,Blaback:2013fca,Blaback:2013qza,Abel:2014xta,Ruehle:2017mzq,Cole:2019enn,Cole:2021nnt,Abel:2021rrj,Loges:2021hvn,Abel:2023zwg}. In particular, they are capable of outperforming deterministic algorithms due to the large size of and computational complexity in the landscape. This is already the case for simple toy models \cite{Denef:2006ad,Bao:2017thx,Denef:2017cxt,Halverson:2018cio}.

In the context of the KS database, GAs have been applied to construct reflexive polytopes in various dimensions \cite{Berglund:2023ztk}. In contrast, applications of GAs or other optimization techniques to polytope triangulations have been rather scarce, see however \cite{Demirtas:2020dbm,Bao:2021ofk,Berman:2021mcw,Berglund:2021ztg} for applications of random walks and Machine Learning. Schematically, the types of constraints that we would like to address with GAs often take the form $g(x) > y$ or $g(x) \approx y$ for some target value $y$. Here, $g$ are functions defined on the space of triangulations of a given polytope. They map UV parameters $x$ like intersection numbers, moduli values or instanton charges to properties of the low energy physics in four dimensions. Consequently, we will focus on \emph{maximization problems} $x = \text{argmax}(g)$ and \emph{inverse problems} $x = g^{-1}(y)$, respectively: that is, can we use GAs to find string compactifications that either maximize low energy parameters or match particular desired values? This is the goal of this work.

As explained in many standard textbooks like \cite{de2010triangulations}, regular triangulations, including FRSTs, are typically represented either by their simplices or as vectors of heights. It turns out, however, that there is a more suitable representation for our purposes. As a result of Wall's theorem \cite{wall}, all FRSTs of a single four-dimensional reflexive polytope with the same induced two-face triangulations give rise to the same Calabi-Yau hypersurface \cite{Demirtas:2020dbm}. For this reason, it is beneficial to use an encoding which avoids such trivial redundancies, see Fig.~\ref{fig:ratioFRSTs} for the count of such redundancies at $h^{1,1}\leq 7$. In doing so, we will work on the level of \emph{two-face inequivalent FRSTs} which we refer to as NTFE (``non-two-face equivalent'') FRSTs.

In this paper, we achieve such a representation by enumerating the fine, regular triangulations of each two-face of the polytope that is being optimized. An FRST can then be encoded by (the indices in the previous enumeration of) its collection of two-face triangulations, which we refer to as its ``DNA''. This circumvents the redundancy of FRSTs with the same two-face restrictions, while still enabling quick conversion to a Calabi-Yau geometry via an \emph{extending} prescription introduced in \cite{macfadden2023efficient}, also discussed in \S\ref{sec:NTFEs}. Importantly, as we explain further below, not any collection of two-face triangulations can be extended to a full FRST of the polytope because ensuring global regularity can be somewhat nuanced. Said differently, even when starting from regular triangulations of each two-face, there may not always be a regular triangulation of the entire polytope with said induced two-face triangulations.

We present three applications in the bulk of this paper. We begin by studying a polytope with Hodge numbers $(h^{1,1},h^{2,1})=(23,7)$ for which we can easily enumerate all NTFE FRSTs. In this example, we introduce and compare two different notions of distance on the space of NTFE FRSTs, namely \emph{Hamming} and \emph{flip distance}, and study the continuity of target functions with respect to these distances. Further, we demonstrate that our GA with optimized hyperparameters easily outperforms standard sampling (like random sampling) and optimization techniques (like Simulated Annealing). As a second application, we investigate axion decay constants for $C_4$-axions in Type IIB compactifications. Specifically, we use our GA to find axions with a specified value $f_*$ for their axion decay constant for a polytope with $(h^{1,1},h^{2,1})=(60,4)$. As the third example, we maximize axion-photon couplings for the largest polytope $(h^{1,1},h^{2,1})=(491,11)$ in the KS database.

The GA developed in this paper, as well as a larger suite of tools for sampling and optimization, are implemented in an open-source Python package \texttt{cyopt} \github{https://github.com/sheride/cyopt}. Sampling and optimization of FRSTs through their DNA parameterization as well through the FRST flip graph are supported through a direct interface with \texttt{CYTools}, but more general search spaces parameterized by integer tuples and graphs are accommodated, as are the additional discrete sampling and optimization methods introduced in \S\ref{sec:h11_23}.

The outline of this paper is as follows. In \S\ref{sec:CYs}, we review the construction of Calabi-Yau threefolds from triangulations of four-dimensional reflexive polytopes. Subsequently, in \S\ref{sec:GAs}, we present our implementation for the GA. Next, we apply our GA to three different polytopes in \S\ref{sec:examples}. We conclude with discussion and outlook in \S\ref{sec:conclusion}.

\section{Calabi-Yau Hypersurfaces from Reflexive Polytopes}\label{sec:CYs}

Let us start by briefly reviewing the construction of Calabi-Yau threefold hypersurfaces in toric varieties.\footnote{More details on this procedure can be found e.g. in \cite{Demirtas:2020dbm,Braun:2017nhi}.} The complete classification of Kreuzer and Skarke leads to 473,800,776 reflexive polytopes in four dimensions \cite{Kreuzer:2000xy}. For each such four-dimensional reflexive polytope $\Delta^\circ$, a fine, regular, star triangulation (FRST) $\cT$ describes a fan for a toric variety $V^{\circ}$ in which the generic anti-canonical hypersurface defines a smooth Calabi-Yau threefold $X^{\circ}$ \cite{Batyrev:1993oya}.

Regular triangulations of polytopes can be represented by a vector of heights which lifts the point collection of $\Delta^\circ$ to one higher dimension, see e.g. \cite{Demirtas:2020dbm}. The lower faces of the convex hull for such a configuration correspond to simplices forming a triangulation of $\Delta^\circ$ upon projection. The space of all height vectors defining a given triangulation of $\Delta^\circ$ defines (the interior of) a polyhedral cone called the \emph{secondary cone}. The collection of secondary cones for all triangulations of $\Delta^\circ$ defines a polyhedral fan called the \emph{secondary fan}. One can similarly define subfans, such as the collection of cones giving rise to FRSTs. The support of the subfan of secondary cones defining all FRSTs is itself a polyhedral cone.

However, as is well known, there is a massive redundancy when going from polytope triangulations to Calabi-Yau hypersurfaces. For this, we recall Wall's theorem \cite{wall} which asserts that Hodge numbers, triple intersection numbers, and second Chern classes completely determine the homotopy type of a compact, simply connected Calabi-Yau threefold with torsion-free homology. For Calabi-Yau hypersurfaces, Hodge numbers are fixed by polytope data, while triple intersection numbers and second Chern classes are determined purely by the induced triangulations of two-faces. Therefore, FRSTs of $\Delta^{\circ}$ with identical restrictions to two-faces give rise to topologically equivalent Calabi-Yau threefolds. When performing optimization on the space of FRSTs, it will be beneficial to find an \emph{encoding} for FRSTs which avoids such trivial redundancies.

\subsection{Generating two-face inequivalent triangulations}\label{sec:NTFEs}

To reiterate, the two-face triangulations of $\Delta^\circ$ define its associated Calabi-Yau threefold (if one exists). We arbitrarily choose an ordering of two-faces, $\Theta^\circ_0, \dots, \Theta^\circ_{n-1}$. For each two-face, we define an arbitrary ordering of its fine, regular triangulations (FRTs) and assign integer labels to each of them. Under these choices, a collection of two-face triangulations can be encoded in a length-$n$ tuple
\begin{equation}\label{eq:dna}
    \cC = (c_0,\dots,c_{n-1})\in \mathbb{N}^{n} \kom 0\leq c_i < N_{\text{FRT}}(\Theta^{\circ}_{i})
\end{equation}
for which the $i$-th entry, $c_i$, is an integer (the label) defining the chosen triangulation of the $i$-th two-face $\Theta^\circ_i$. Further, $N_{\text{FRT}}(\Theta^{\circ}_i)$ denotes the number of FRTs of the two-face $\Theta^\circ_i$. Assuming one finds a map from $\cC$ to an FRST $\cT$ of $\Delta^\circ$, the advantage of this encoding is that different integer vectors $\cC$ define two-face inequivalent FRSTs which we refer to as \emph{NTFE FRSTs}. In the GA-context below, we call choices $\cC$ of two-face triangulations as the \emph{DNA} or \emph{chromosome} of the associated Calabi-Yau hypersurface. 

For such an encoding to be most useful, one needs an efficient method to map DNA $\cC$ to the associated FRSTs $\cT$ of $\Delta^\circ$. In practice, the algorithm developed by one of the authors \cite{macfadden2023efficient} achieves precisely that. Specifically, instead of constructing the full set of FRSTs first and then modding out by two-face inequivalence,\footnote{For $h^{1,1}\geq 10$, this is usually computationally infeasible and even at lower values of $h^{1,1}$ requires huge memory resources. For instance, computing all FRSTs for the unique polytope with $(h^{1,1},h^{1,2})=(9,19)$ required more than several hundreds of GBs of memory leading to $\#\text{FRSTs}= 162917$, while directly generating the only 14 NTFE FRSTs using the methods described here required $4$MB.} one can efficiently construct a polyhedral cone for which any vector in its strict interior generates the desired two-face restrictions $\cC$. To be more precise, this cone is the intersection of the secondary cone of each two-face triangulation, embedded in the height space of the entire polytope. Any vector in the strict interior of this intersection is also in the strict interior of each two-face cone. Thus, it defines a triangulation $\cT$ with the imposed two-face restrictions, $\cC$. In this way, two-face inequivalent triangulations are generated much more time and memory efficiently by requiring orders of magnitude fewer operations.

For our purposes, we need to understand which choices of two-face triangulations $\cC$ are realizable as a regular triangulation $\cT$ of the ambient polytope. This is equivalent to asking whether the intersection of the secondary cones is full-dimensional. This is achieved by constructing the constraint programming problem of finding any point, $x$, satisfying $Hx>0$, where $H$ are the inwards-facing hyperplane normals of the intersected cone. If no such $x$ exists, then no triangulation of the polytope has said two-face restrictions $\cC$.\footnote{We recall that only height vectors in the strict interior of the secondary cone give rise to the desired triangulation.}

In this way, one can directly generate an FRST\footnote{Strictly, the produced triangulation is only guaranteed to be fine with respect to two-faces. This is acceptable because the anti-canonical divisor defining the Calabi-Yau hypersurface does not intersect the divisors associated with points interior to facets. Additionally, while the produced triangulation may not be star, it can trivially be made star (without affecting the two-face triangulations) by lowering the height of the origin.} from given two-face triangulations, or prove that no such FRST exists. The process $\cC\rightarrow \cT$ of mapping two-face triangulations $\cC$ to this cone, finding an interior point, and generating the corresponding FRST $\cT$ is called \emph{extension}. This extension can fail whenever the cones have a non-solid intersection. Thus, there will be no FRST with said two-face restrictions. We will see this explicitly in our examples below, see in particular Fig.~\ref{fig:h1160_failedlifts}.\footnote{Without giving away too much (see \S\ref{sec:GAimplementation} for details), we discard such non-extendable DNA when running the GA.}

In the above encoding, the relevant enumeration, then, is of fine, regular triangulations (FRTs) of the two-faces $\Theta^{\circ}$ of $\Delta^\circ$. The number of such FRTs $N_{\text{FRT}}(\Theta^{\circ})$ of a two-face $\Theta^{\circ}$ is usually modest since the two-faces live in a lower dimension (i.e., 2D), have typically fewer points, etc. than $\Delta^\circ$. For example, the first $1000$ polytopes with $h^{1,1}=20$ in the ordering of KS each have less than $53$ two-faces with each two-face having less than $169$ FRTs. Further, even at high $h^{1,1}$, it is not rare for two-faces to be trivial, having only one FRT. Typically, a good measure for the complexity of a given two-face $\Theta^{\circ}$ is the number of points which we denote by $\ell(\Theta^{\circ})$.

To summarize, all that is needed to encode the FRST is a modest-length vector (length $\sim \mathcal{O}(10-100)$) of typically modest-sized entries, indicating the two-face triangulations defining the FRST. The extending procedure then efficiently maps these indices into a Calabi-Yau manifold.

\subsection{Equivalences beyond two-face triangulations}\label{sec:equivalence}

Before we continue, let us briefly comment on the role of NTFE FRSTs within the context of diffeomorphic Calabi-Yau threefolds. In addition to two-face equivalence, automorphisms leaving $\Delta^{\circ}$ invariant produce additional redundancies on the level of triangulations since these lead to trivial identifications of the associated Calabi-Yau threefolds. In \cite{Gendler:2023ujl}, \emph{FRST classes} were introduced as those sets of FRSTs having identical restrictions to two-faces, up to the action of an automorphism of the polytope. 

Beyond that, the question of topological inequivalence is highly non-trivial since it requires finding an integral change of basis $\L\in\operatorname{GL}(h^{1,1},\bbZ)$ between the triple intersection forms and second Chern classes $(\kappa_{ijk},c_{2,i})$ and $({\kappa^\prime}_{ijk}, {c^\prime}_{2,i})$ of two compact Calabi-Yau threefolds $X$ and ${X^\prime}$, respectively. Recently, \cite{Gendler:2023ujl} proved the exact number of distinct simply connected Calabi-Yau threefold hypersurfaces at $h^{1,1}\leq 5$ obtained from triangulations of four-dimensional reflexive polytopes, while obtaining stringent bounds for $h^{1,1}=6,7$.\footnote{Independently, the authors of \cite{Chandra:2023afu} obtained bounds consistent with this analysis at $h^{1,1}\leq 6$.} Incorporating such equivalences in an optimization framework would be highly desirable, but in the absence of a systematic approach at higher values of $h^{1,1}$, this is beyond the scope of this work.

\begin{figure}[t!]
  \centering
  \includegraphics[width = \linewidth]{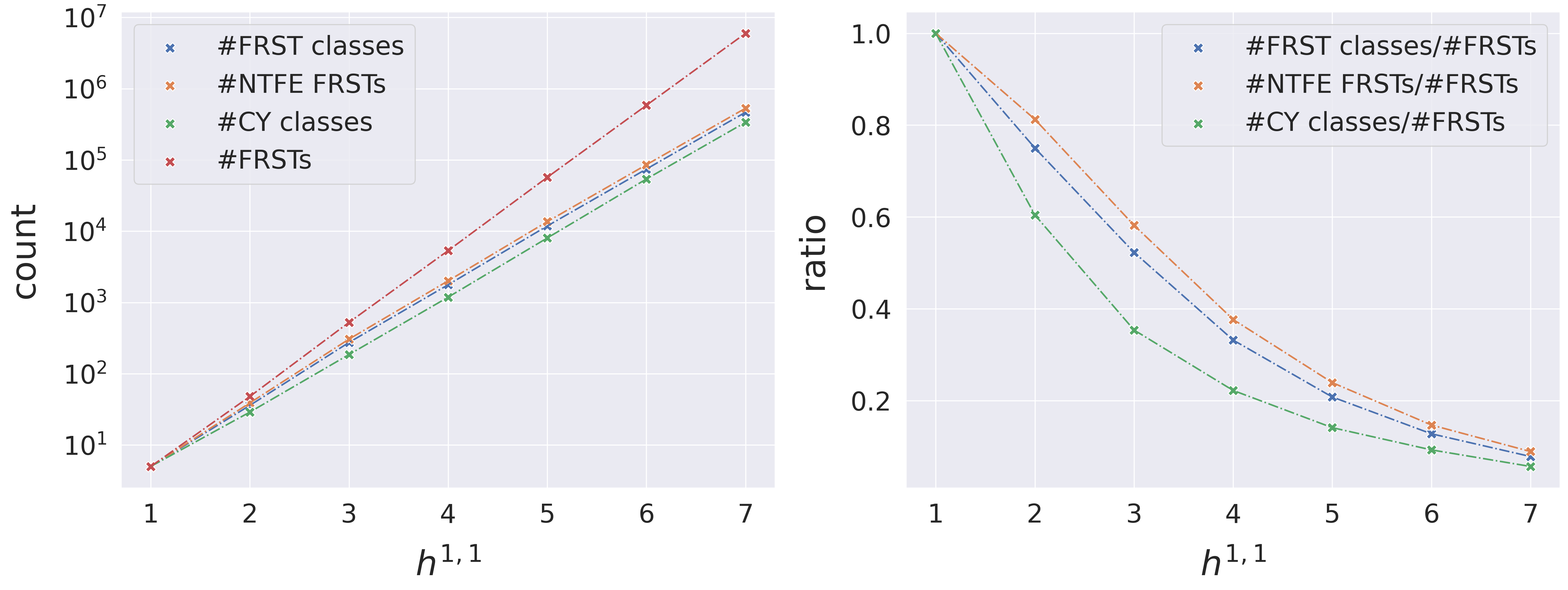}
  \caption{\emph{Left}: Total number of NTFE FRSTs, FRST classes, Calabi-Yau (CY) classes and FRSTs for all polytopes at $h^{1,1}\leq 7$. \emph{Right}: Ratio of different equivalence classes of FRSTs over the number of FRSTs.}\label{fig:ratioFRSTs}
\end{figure}

However, let us argue that, with the encoding defined above, we are able to account for a large fraction of redundancies already. In Fig.~\ref{fig:ratioFRSTs}, we show the ratio of NTFE FRSTs and total number of FRSTs for all polytopes at $h^{1,1}\leq 7$. We observe an exponential falloff of the number of NTFE FRSTs relative to that of FRSTs. We also included the equivalent ratios for FRST classes and Calabi-Yau classes, respectively, for all polytopes at $h^{1,1}\leq 7$ in Fig.~\ref{fig:ratioFRSTs}.\footnote{We use the lower bounds for the number of Calabi-Yau classes at $h^{1,1}=6,7$ obtained in \cite{Gendler:2023ujl}.} Out of all FRSTs at $h^{1,1}\leq 7$, only $9.55\%$ define NTFE FRSTs, $8.37\%$ FRST classes and $6.04\%$ Calabi-Yau classes \cite{Gendler:2023ujl}. 
Extrapolating to higher $h^{1,1}$, while the trend for CY classes remains unknown, \cite{Demirtas:2020dbm} derived bounds on the number of FRSTs as well as NTFE FRSTs\footnote{It is expected that automorphism equivalence only amounts to an $\mathcal{O}(1)$ fraction \cite{Demirtas:2020dbm}. We checked that this is indeed the case in all of our examples discussed below.} across the entire KS database
\begin{equation}\label{eq:bounds_KS_landscape}
    \#\text{FRSTs}\lesssim 1.53\times 10^{928} \kom \#\text{NTFE FRSTs}\lesssim 1.65 \times 10^{428}\, .
\end{equation}
The main takeaway is that there is an exponential reduction from \#FRSTs to \#NTFE FRSTs, thus supporting our choice of encoding.

We deduce that a significant redundancy when mapping FRSTs to Calabi-Yau geometries can be attributed to trivial equivalences defining NTFE FRSTs. With our encoding, these redundancies are automatically removed, thereby ensuring that our optimization strategies are less prone to encountering equivalent Calabi-Yau threefolds.

\section{Genetic Algorithms for Polytope Triangulations}\label{sec:GAs}

In this section, we describe our implementation for a GA for FRSTs of four-dimensional polytopes. While there are many publicly available packages for GAs such as \cite{1995ApJS..101..309C,charbonneau1995user,charbonneau2002introduction,charbonneau2002release},\footnote{We refer to \href{http://geneticprogramming.com/software/}{http://geneticprogramming.com/software/} for a maintained list of packages for genetic programming in general.} we implemented our own GA which is specialized to polytope triangulations.

\subsection{Genetic Algorithms}

\begin{table}[t!]
  \centering
  \begin{tabular}{c|c}
  & \\[-1em]
  Genetic Algorithms & Kreuzer-Skarke database \\ [0.1em]
  \hline 
  \hline 
  & \\[-1em]
  individuals & Calabi-Yau hypersurfaces\\ [0.1em]
  \hline 
  & \\[-1em]
  chromosome & two-face FRT IDs $\cC$, cf.~Eq.~\eqref{eq:dna}\\ [0.1em]
  \hline 
  & \\[-1em]
  alleles & two-face FRT ID $c_i$ \\  [0.1em]
  \hline 
  & \\[-1em]
   phenotype & volume, axion decay constants etc. \\ [0.1em]
  \hline 
  & \\[-1em]
  boundary conditions & extending of two-face FRTs to FRST \\ [0.1em]
  \end{tabular} 
  \caption{Dictionary relating GA terms to polytope NTFE FRSTs.}\label{tab:GADic} 
\end{table}

With \texttt{CYTools} \cite{Demirtas:2022hqf}, it is straightforward to construct FRSTs for any given polytope in the KS database within seconds. However, the far more interesting task is obtaining such FRSTs with special properties such as large Kähler metric eigenvalues. \emph{Genetic Algorithms}\index{Genetic Algorithms} (GAs) \cite{Holland1975,David1989,Holland1992,reeves2002genetic,haupt} are perfectly suited to address this problem and to scan the landscape of Calabi-Yau hypersurfaces for desired features. They are based on processes inspired by natural evolution and designed to search huge parameter spaces for optimal solutions.

We use the dictionary shown in Tab.~\ref{tab:GADic} to relate the GA terminology to polytope triangulations. As usual, we start from a random population of individuals which in our case correspond to Calabi-Yau manifolds associated with FRSTs $\cT$ of a polytope $\Delta^\circ$. More specifically, the members of a population are NTFE FRSTs describing homotopy inequivalent Calabi-Yau geometries. Each individual is characterized by its genotype and phenotype. The former is a genetic code or \emph{chromosome} $\cC$, i.e., the list of two-face triangulation IDs in \eqref{eq:dna}, built from parameters in the input space. The latter defines its properties in the output space such as the Calabi-Yau volume $\mathcal{V}$ or axion decay constants $f$, collectively denoted by $\mathcal{O}$ in the following. For many applications, we want to find configurations in which such observables take particular value, namely $\mathcal{O}\approx \mathcal{O}_*$ for some target value $\mathcal{O}_*$. To achieve this, the population evolves over several generations as dictated by a fitness function $F(\mathcal{O},\mathcal{O}_*)$ defined on the phenotype space. The fitness function $F$ peaks around the desired value $\mathcal{O}_*$ and is typically chosen as a Gaussian or rational function.\footnote{We will comment on these choices for the individual examples in \S\ref{sec:examples}. Let us also note that generally the performance of the GA is sensitive to the choice of fitness function.} Based on Darwin's principle of survival of the fittest with each generation, the population maximizes the fitness, thereby more and more approximating the desired properties in its phenotype.

Crucially, such a GA would be able to explore the secondary subfan of all NTFE FRSTs \emph{non-locally}. In our implementation, a single crossover or mutation operation can affect multiple triangulations of two-faces at once, thereby jumping across many chambers of the secondary subfan irrespective of their relative distance. This needs to be distinguished e.g. from flop transitions which take us in neighbouring phases by traversing walls of the extended Kähler cone. 

As argued e.g. in \cite{Cole:2019enn} in the context of Type IIB flux vacua, GAs are more powerful in finding new physical solutions to the $F$-term equations than e.g. random sampling of flux vectors. Indeed, they detect underlying structures in the flux landscape that are typically unknown at the outset of the search. In the present setting, one would expect the GA to be more efficient in locating configurations $\cC$ of two-face triangulations (recall Eq.~\eqref{eq:dna}) which extend consistently to FRSTs $\cT$ of the four-dimensional polytope $\Delta^\circ$ than just random samples of two-face triangulations. Indeed, we observe such a behaviour in our results below, cf.~Fig.~\ref{fig:h1160_failedlifts}.

Despite these successes, there are also limitations to GAs or stochastic search algorithms in general. They perform just as poor as random scans when trying to solve ``needle-in-a-haystack'' type problems. This happens frequently in highly tuned regimes where the phenotype space becomes too sparse or, said differently, the associated distributions develop tails. Vice versa, an excellent performance is expected whenever there exists a pseudo-continuous neighbourhood around the optimal solution which might be characterized by a funnel-like topography (see e.g. \cite{bryngelson1995funnels,Khoury:2019yoo}). Below, by exploring different notions of distance between NTFE FRSTs, we argue that such structures are indeed present in the energy landscapes of polytope FRSTs.

\subsection{Implementation for polytope triangulations}\label{sec:GAimplementation}

Let us now describe the implementation of our GA in more detail.\footnote{Our implementation is based on a version of the GA described in \cite{Schachner:2022bak} for Type IIB flux vacua.} Concretely, we begin with a set of NTFE FRSTs defined by their DNA $\lbrace\cC_{1},\ldots,\cC_{P}\rbrace$, called a population of size $P$, with $\cC_{i}\in \mathbb{N}^{n}$ as defined in \eqref{eq:dna} where $n$ denotes the number of two-faces of the polytope $\Delta^\circ$. We reiterate that, in our encoding, the DNA $\cC = (c_0,\ldots ,c_{n-1})$ associated with a Calabi-Yau manifold is represented by the triangulation labels $c_i$ for each two-face $\Theta^{\circ}_i$. In particular, this means that each position $c_i$ in $\cC$ has a fixed range determined by the maximal number of available FRTs of $\Theta^{\circ}_i$. Then, the algorithm involves the following four steps:
\begin{enumerate}[label=(\Alph*)]
    \item The GA selects individuals based on some probability distribution obtained from the fitness through suitable selection principles to be defined below.
    \item The selected representatives are subsequently used to construct new FRSTs $\tilde{\cC}_{l}$ by performing crossover operations acting on pairs of DNA $(\cC_{i},\cC_{j})$.
    \item The third (and arguably most important) step concerns mutation, which alters the resulting vectors $\tilde{\cC}_{l}$ by some randomized procedure.
    \item Lastly, one defines a novel population through a survival description and repeats the above process. Each iteration is referred to as \emph{generation} and the maximal number of generations will henceforth be denoted $G$.
\end{enumerate}
Each individual step described above involves a choice of operations of how Calabi-Yau manifolds are selected, their genetic information is exchanged and randomly altered, see e.g.~\cite{davis1991handbook,Ruehle:2020jrk} for a comprehensive list.

Specifically, our implementation utilizes the following operators:
\begin{enumerate}
\item \textbf{Selection:} The fittest individuals should be more likely to procreate.
There are several ways to select Calabi-Yau manifolds based on the predefined fitness:
\begin{itemize}
\item \emph{roulette wheel selection} (RWS): we take the normalized fitness as a probability distribution to draw a given number of samples from the population.
\item \emph{tournament selection} (TS): we choose $k$ random individuals which compete against each other in a tournament.
The fittest individual wins and is selected.
\item \emph{rank selection} (RS): we assign ranks $r$ to each individual based on their fitness from which define again a linear probability distribution $1-r/P$ to select individuals for. In this way, a more balanced selection process can be ensured.
\item \emph{truncated rank selection} (TRS): this is similar to RS, but only a certain percentage of fittest individuals takes part in the selection process.
\item \emph{exponential rank selection} (ERS): this is also similar to RS, but the probability distribution is an exponential function of the ranks (e.g. $1-\text{e}^{r-P}$) instead of a linear one.
\end{itemize}
It will turn out to be helpful to work with several selection principles to avoid getting struck in local minima. In particular, RWS leads typically to hierarchical probabilities, while TS or RS have more evenly spaced distributions.
\item \textbf{Crossover:} From the selected set of Calabi-Yau manifolds, we build pairs of chromosomes $(\cC_{i},\cC_{j})$ on which we perform crossover operations, thereby constructing a new chromosome $\tilde{\cC}$. Again, we distinguish several different options summarized
\begin{itemize}
\item \emph{single point crossover} (SPX): we cut both DNA of $\cC_{i},\cC_{j}$ at the same random locus $\lambda$. Then $\tilde{\cC}$ receives the alleles $1$ to $\lambda$ from $\cC_{i}$ and $\lambda+1$ to $n$ from $\cC_{j}$.
\item \emph{$k$-point crossover} (kPX): we cut the chromosomes of $\cC_{i},\cC_{j}$ at $k$ random loci $\lambda_{l}$, $l=1,\ldots,k$. Then $\tilde{\cC}$ is constructed by splicing the different pieces from $\cC_{i}$ and $\cC_{j}$ together in an alternating way.
\item \emph{uniform crossover} (UX): for each allele, we make a random choice whether $\tilde{\cC}$ inherits the allele from $\cC_{i}$ or $\cC_{j}$.
\end{itemize}
\item \textbf{Mutation:} Random modifications of the new chromosomes are essential to avoid stagnation.
Mutation is however not simply a tool to improve convergence,
but rather an integral feature of GAs allowing them to explore previously unknown areas of the solutions space.
There are various ways in which this can be implemented:
\begin{itemize}
\item \emph{random mutation} (RM): we select $k$ random loci and replace the two-face triangulation IDs by a uniformly random integer in the required range. Here, the range is determined by the input of two-face triangulations. We are mostly using $k = 1, 2$ in our applications below.
\end{itemize}
In addition to these operators, we include a global \emph{mutation rate} $r_{M}$ which sets the total amount of mutations performed on the complete dataset. This is necessary because too much stochasticity undermines the GA’s ability to find patterns associated with the optimal solution, see e.g. \cite{Cole:2019enn}. Typically, we set $0.01 \lesssim r_{M} \lesssim 0.1$ so that $1-10\%$ of individuals are being mutated.
\item \textbf{Population generation:} After constructing the new two-face triangulations, we check whether they extend to a full FRST of the polytope $\Delta^\circ$ as outlined above (recall \S\ref{sec:NTFEs}). We discard those DNA which fail to extend to an FRST, while the remaining ones are used to construct a new population. In particular, we achieve this by assigning a target of $-\infty$ to such DNA, which precludes their selection in the GA. We additionally enforce that DNA are pairwise distinct in the new population.
\item \textbf{Survival:} To ensure the stability of the algorithm, it is helpful to employ \emph{fitness-based survival} methods. That is, after we constructed a population of $P$ new solutions in the previous step, we replace the least fit individuals by fitter ones from the previous generation/population referred to as \emph{survival of the fittest} (see e.g. \cite{Ruehle:2020jrk}). This ensures that a fixed number of best solutions is always carried over to the next generation such that the overall fitness can only increase. 
\end{enumerate}

A few comments are in order. First, in many applications, only a small number of operators are necessary for a successful search. These are controlled by so-called \emph{hyperparameters} specifying the rate with which a certain operator is being applied. In total, we have $\sim 14$ hyperparameters, namely
\begin{enumerate}[label=(\roman*)]
    \item 6 for selection (4 to parameterize relative frequency of 5 selection types, the tournament size, and the TRS percentage)
    \item 2 for crossover (to parameterize relative frequency of 3 crossover types: SPX, kPX for $k = 2$, and UX)
    \item 2 for mutation (1 to parameterize relative frequency of 2 mutation types --- RM for $k = 1, 2$ --- and the mutation rate $r_M$)
    \item 1 for survival (number of survivors)
\end{enumerate}
along with the total population size and fitness function hyperparameters: as we will discuss, we typically adopt a inverse square or Gaussian fitness, which have $1$ and $2$ hyperparameters, respectively. Crucially, we do not pick the hyperparameters by hand, but we use additional optimization techniques to tune them for a given task, see \S\ref{sec:hyperopt} blow.

Second, let us comment on the enumeration of two-face triangulations. Depending on the scope, our GA runs with different modes for generating said two-face triangulations:
\begin{itemize}
\item $\mathtt{mode=all}$ uses all FRTs of all two-faces which is generically feasible if $\ell(\Theta^{\circ}_i)\leq 17$ for all two-faces $\Theta^{\circ}_i$. This mode is not really limited by $h^{1,1}$, but the configurations of two-faces.
\item $\mathtt{mode=random}$ samples FRTs randomly for some two-faces which becomes necessary if $\ell(\Theta^{\circ}_i)> 17$. For the random sampling, we use methods described in \S5 of \cite{Demirtas:2020dbm}.
\item $\mathtt{mode=file}$ picks FRTs of two-faces from data files. This allows us to run the GA on the same sets of two-face triangulations.
\end{itemize}

\subsection{Hyperparameter tuning from Bayesian optimization}\label{sec:hyperopt}

As mentioned above, our GA implementation requires a choice for $\sim 14$ hyperparameters. Instead of fixing those parameters by hand, we optimize them using \emph{Bayesian optimization} (BO) \cite{rasmussen2005gaussian}.\footnote{The use of BO to optimize hyperparameters for evolutionary algorithms has been explored previously in e.g. \cite{roman2016bayesian,karro2017black,ruther2021bayesian}. For a broader overview over hyperparameter tuning, we refer to the reviews \cite{smit2009comparing,huang2019survey}.} This optimization algorithm is computationally expensive but requires no gradient data, is resistant to local minima, and is designed to make the most of every point sampled: thus, it is ideal for global optimization problems on expensive functions \cite{frazier2018tutorial, garnett_bayesoptbook_2023}. We want to maximize the performance of a GA, averaged over many runs, as a function of its hyperparameters, a generically expensive function.

Bayesian optimization extremizes a function $f$ by modeling its functional values as random variables and using Bayesian inference to both predict its values away from sampled points and choose the best points to sample next\footnote{This paragraph assumes some basic familiarity with Bayesian inference and (Gaussian) processes: for reviews, see \cite{RevModPhys.83.943}.}. In particular, $f$ is modeled as a stochastic process, or a sequence of random variables $g(x)$, one for each point $x$ in the domain. Often --- and in our case --- $f$ is modeled as a Gaussian process in particular: this means for any finite set of points $x_1, \dots, x_n$, the random variables $g(x_1), \dots, g(x_n)$ are distributed according to a multivariate Gaussian. That is, our random variables $g(x)$ are parameterized by functions $\mu(x)$ and $\Sigma(x,x')$ which tell us the mean of $g(x)$ and the covariance between $g(x)$ and $g(x')$, respectively. 

As with all Bayesian inference problems, we must begin by selecting a prior on our theory parameters $\mu(x)$ and $\Sigma(x,x')$: typically, we set $\mu_0(x) = 0$ and $\Sigma_0(x,x')$ to be some monotonically increasing function of $\|x - x'\|$, ensuring nearby points in the domain are correlated (this captures the continuity of $f$). From here, we assume that we have sampled $f$ at $x_1, \dots, x_n$, and we endeavor to use this information to compute a posterior on $g(x)$. This can be thought of in a few ways. First, we can think about conditioning $g(x)$ on the requirement that $g(x_i) = f(x_i)$, $1 \leq i \leq n$ (recalling that conditioning a multivariate Gaussian on certain values yields another, lower dimensional Gaussian). Equivalently, we can think about using Bayes' theorem to compute the posterior on the process $g(x)$ by multiplying the aforementioned prior with the likelihood $\mathcal{L} = \prod_i \delta(g(x_i) - f(x_i))\cdot \delta(\Sigma(x_i, x_i))$. Either way, this provides us with a new, updated distribution for the process $g$, informed by existing samples, with the means $\mu(x)$ being the expected values for $f(x)$ and the standard deviations $\Sigma(x,x)$ capturing the model's uncertainty in that expected value. More specifically, as more points are sampled, the means $\mu(x)$ will approach $f(x)$ and the standard deviations $\Sigma(x,x)$ will shrink.

Beyond modeling $f(x)$ as the means of the posterior distribution of $g(x)$, this statistical approach sheds light on the optimal points to sample next: in general, we want to balance exploitation (sampling near previously observed) and exploration (sampling points far from previously sampled points), and the means of our process reveal where the function is largest (where we ought to exploit) while the standard deviations establish where we understand $f$ the least (where we ought to explore). In practice, a user intructs a Bayesian Optimization algorithm to sample $N$ points in the domain: the first $M$ are sampled randomly and then the subsequent $N - M$ are chosen by maximizing\footnote{Let us stress that acquisition functions are nicer and quicker to evaluate than the expensive $f$, and are thus much more amenable to simpler optimization methods.} so-called acquisition functions, such as the expected improvement of the largest functional value sampled thus far. For further discussion on acquisition functions, see e.g. figures $1.2-1.4$ and the surrounding text in \cite{garnett_bayesoptbook_2023}. Upon evaluating $f$ at these $x_1, \dots, x_N$, the algorithm then returns $\mathrm{argmax}_{x_i} f$, the $x_i$ corresponding to the largest seen value of $f$.

In our case, the expensive function $f$ maps the space of hyperparameters (the domain) to the average best target across $m$ runs of the GA, where a run terminates once it has seen $k$ unique DNA: i.e., it has had to evaluate the expensive target function $k$ times. This computational cost varies between our use-cases and motivated the choice of different $m, k$ values in each case. For example, for our $h^{1,1} = 23$ application of \S\ref{sec:h11_23}, we chose $m \sim 50$ and $k \sim 200$, as we found model performance to be characterized largely by performance within the first $200$ DNA. In our implementation, we used the \texttt{BayesianOptimization} python package \cite{bayesopt}, and found success when using its Sequential Domain Reduction functionality.

\section{Applications to String Compactifications}\label{sec:examples}

In this section, we present three applications of the GA described in the previous section. We begin by maximizing the Calabi-Yau volume for a polytope with modest $h^{1,1}=23$ for which all NTFE FRSTs can be easily obtained. This allows us to demonstrate the GA's ability to locate the fitness maximum and to quantify its performance. Subsequently, we pick a polytope at $h^{1,1}=60$ for which all two-face FRTs can be easily computed. Here, we utilize the GA to find solutions with a specific value for the decay constant $f$ of the lightest $C_4$ axion in compactifications of Type IIB string theory. Lastly, we run our GA to maximize axion-photon couplings as computed in \cite{Gendler:2023kjt} for the largest polytope in the KS database with $h^{1,1}=491$.

\subsection{Maximizing the volume --- $h^{1,1}=23$}\label{sec:h11_23}

We begin by applying the GA to a maximization problem for a polytope $\Delta^\circ$ with $h^{1,1} = 23$. In particular, this polytope is sufficiently simple that we can compute the full set of NTFE triangulations and their associated features: that is, we can map out the entire search space, making this a nice testing ground for the GA, even if the brute force approach is technically feasible. In particular, we can \emph{theoretically} examine whether the distribution of features near that maximum has the ``funnel-like topography'' which GAs are equipped to exploit and then \emph{empirically} assess how effective the GA is in finding the true global maximum (especially in comparison to other methods).

The vertices of $\Delta^\circ$ read
\begin{equation}
    \begin{pmatrix}
        1 & 0 & 0 & 0 & 0 & 2 & -2 & -1 & 0 & 1 \\
        0 & 1 & 0 & 0 & 0 & 2 & -1 & -2 & 1 & 0 \\
        0 & 0 & 1 & -1 & 1 & -1 & 0 & 2 & 0 & -2 \\
        0 & 0 & 0 & 0 & 2 & -2 & 2 & 2 & -2 & -2 
    \end{pmatrix} \, .
\end{equation}
This polytope has 8 two-faces with more than one FRT, specifically 4 two-faces with either 4 or 6 FRTs respectively. Therefore, a choice of DNA $\cC$ of a Calabi-Yau threefold in this example corresponds to an integer vector of length 8, i.e.,\footnote{The polytope $\Delta^\circ$ has in total 26 two-faces of which 18 have exactly one FRT making them trivial from our point of view. For this reason, we typically omit them in the definition \eqref{eq:DNA_h11_23} of the DNA.}
\begin{equation}\label{eq:DNA_h11_23}
    \cC = (c_0,c_1,\ldots ,c_7)\in [0,\ldots, 3]^4\times [0,\ldots, 5]^4\, .
\end{equation}
This amounts to $4^4 \times 6^4 = 331776$ sets of two-face FRTs, i.e., choices of DNA. By direct computation, we verified that the number of NTFE FRSTs is
\begin{equation}
    \#\text{NTFE FRSTs}= 331192\leq 331776\, .
\end{equation}
Hence, only $584$ combinations of two-face FRTs fail to extend to FRSTs of $\Delta^\circ$.\footnote{The rate of failed extensions varies significantly across polytopes, see in particular Fig.~\ref{fig:h1160_failedlifts} below. The observed rate for this polytope, $584/331776\approx 1.76\times 10^{-3}$, is relatively small.}

Let us start with a simple task to showcase the performance of our GA. We pick the volume $\mathcal{V}$ of the Calabi-Yau threefold $X^\circ$ as our optimization target: the task will be to maximize it. To compute it, we choose a basis $\{\omega_i\}_{i=1}^{h^{1,1}(X^\circ)}$ of $H^2(X^\circ,\mathbb{Z})$ and define the K\"ahler class $J=\sum_i t^i\,\omega_i$ of $X^\circ$ in terms of K\"ahler parameters $\{t^i\}_{i=1}^{h^{1,1}(X^\circ)}$. The volume of $X^\circ$ is then given by 
\begin{equation}
    \mathcal{V} = \int_{X^\circ} J\wedge J\wedge J= \dfrac{1}{6}\kappa_{ijk} t^i t^j t^k\kom \kappa_{ijk}\coloneqq \int_{X^\circ} \omega_i\wedge \omega_j\wedge \omega_k
\end{equation}
in terms of the K\"ahler parameters $t^i$ and the triple intersection numbers $\kappa_{ijk}$ on $X^\circ$. For convenience, we compute $\mathcal{V}$ at the tip of the stretched K\"ahler cone which is the point closest to the origin in the region of moduli space where the curvature expansion of string theory is well-controlled.

\begin{figure}[t!]
    \centering
    \includegraphics[width=0.95\linewidth]{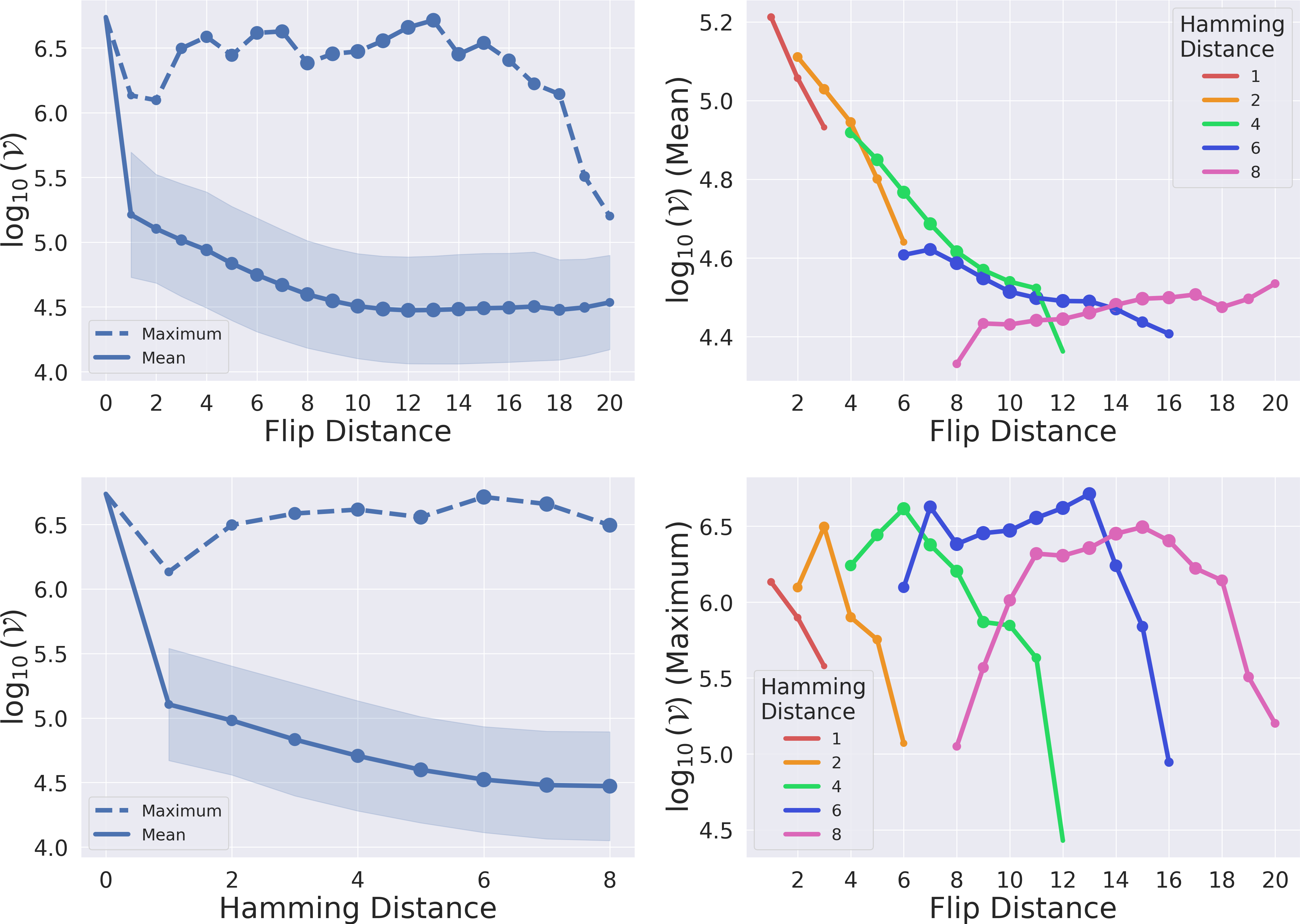}
    \caption{Distance Comparison. \emph{Left:} Mean and maximum Calabi-Yau volume as a function of flip (top) and Hamming (bottom) distance. The fact that the mean target function value correlates with distance supports the idea that the search space of DNA exhibits a funnel-like topography near the global maximum, motivating the use of a GA. \emph{Right:} Calabi-Yau volume as a function of flip distance for fixed Hamming distance. We see that both forms of distance experience correlation with the average behavior of target function when the other is held fixed, suggesting that the target function experience ensemble-level continuity with both Hamming and flip distance independently.
    }
    \label{fig:dist-h11-23}
\end{figure}

First, before we use the GA to explore the space of NTFE FRSTs, it is insightful to scrutinize the structure of features in this space by introducing different notions of distance. One candidate is the \emph{Hamming distance}, defined here as the number of differing two-face triangulations between two FRSTs. This distance is natural in the context of GAs, but adopting the perspective of the secondary fan gives rise to an alternative notion of distance: namely, the \emph{flip distance}, or the minimum number of bistellar flips one must perform to turn one FRST into the other. Equivalently, this is the minimum of walls between distinct secondary cones one must pass through to connect the secondary cones of two FRSTs.

A priori, it is unclear whether features --- $\mathcal{V}$ or otherwise --- will be in any sense ``continuous'' with respect to either of these distances: that is, whether the difference in feature between FRSTs will correlate with either of the two notions of distance. We test this in Fig.~\ref{fig:dist-h11-23}, plotting our feature $\mathcal{V}$ against both notions of distance with respect to the FRST associated with the global maximum. On the left, we examine each distance type separately, and see that FRSTs further away from the global maximum have, on average, smaller feature values. However, features near the value of the global maximum are still achieved at large distances because the decrease in feature mean is overcome by an increase in statistics: there are simply more points at further distances. On the right, we look at the two distance types concurrently, keeping flip distance on the $x$-axis and separating out Hamming distance by color. We can see that for fixed, small Hamming distance, the feature correlates strongly with flip distance, whereas for intermediate ($\approx 10$) flip distance, the feature correlates with Hamming distance.\footnote{We comment that while it could certainly be a fluke that $\mathcal{V}$ behaves roughly continuously with Hamming/flip distance, we found similar behavior with two additional independent features: the number of non-zero intersection numbers and the second-largest GKZ vector component. This suggests that generic features are roughly continuous with respect to these distance measures.} To summarize, we find that the global fitness maximum is indeed surrounded by a funnel-like region. As we will see below, this is a prerequisite for the effectiveness of the GA.

\begin{figure}
    \centering
    \includegraphics[width=0.95\linewidth]{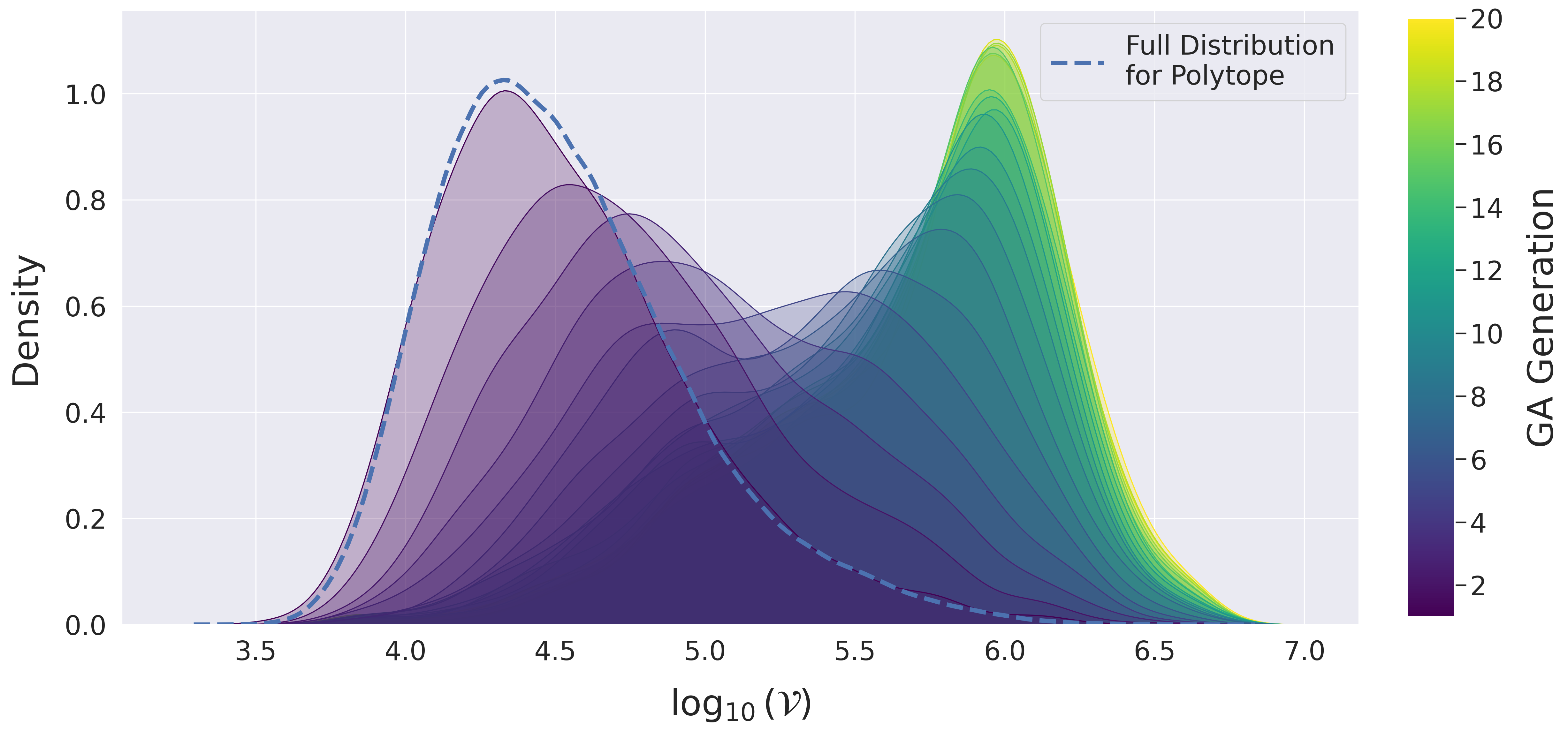}
    \caption{Distribution of values for each generation averaged over $25$ runs for a population size of 100 for the same choice of optimized hyperparameters, but random initial populations. The dashed line shows the full distribution of $\mathcal{V}$ for all NTFE FRSTs of $\Delta^\circ$.
    }
    \label{fig:models-h11-23-resGA}
\end{figure}

With that in mind, we first run our GA with a population size $P = 10$ until each has evaluated $1000$ unique DNA a total of $150$ times.\footnote{A disadvantage of selecting a population size as small as $P = 10$ is that many generations are necessary to explore the search space, as each individual generation introduces few novel DNA (though clock time remains low because previously seen DNA are cached and are thus easily evaluated). We have found some preliminary evidence that dynamic population sizes --- i.e., setting $P > 10$ later in the optimization process --- can ameliorate this problem and improve performance, but we defer a systematic study of this approach to later work.} We note that the global maximum of $\mathcal{V}$ is $\log_{10}(\mathcal{V}_{\mathrm{max}}) \approx 6.73$. For the fitness, we use
\begin{equation}\label{eq:fitness_h11_23}
    F(\mathcal{V},\mathcal{V}_*) = \dfrac{1}{(\log_{10}(\mathcal{V})-\log_{10}(\mathcal{V}_*))^2}\, ,
\end{equation}
where $\mathcal{V}_*$ is a hyperparameter to be fixed momentarily. We run the Bayesian optimization procedure to tune the hyperparameters as described in \S\ref{sec:hyperopt}. Specifically, we performed BO on the GA hyperparameters by evaluating the performance at $\sim 10^4$ different hyperparameter configurations. Noteworthy features of this process include the distinction of a relatively low mutation rate ($\sim 0.05$), strong preferences for tournament selection and one-point random mutation by a factor of five versus other selection/mutation methods, and a tournament size of $4$ (relatively large compared to the total population size $P = 10$). Further, the fitness function \eqref{eq:fitness_h11_23} is determined by $\log_{10}(\mathcal{V}_*)\approx 7.87$.

Initially, we show the evolution of the GA's population in Fig.~\ref{fig:models-h11-23-resGA} colored by generation where we averaged over $25$ individual GA runs with randomly initialized populations. To increase statistics for the purpose of visualization, we choose a population size of $P = 100$, but the GA can learn quicker with fewer unique target function evaluations when we adopt the $P = 10$ population size. In addition, we also show the full distribution for $\log_{10}(\mathcal{V})$ for the entire dataset of $331192$ NTFE FRSTs. This provides a visualization of the GA's ``learning'' process. In particular, after a random population initialization, the GA explores the landscape, quickly finding DNA in the tails of the $\log_{10}(\mathcal{V})$ distribution in the first few generations. Within the first 20 generations, then, the GA is able to effectively construct entire populations with anomalously large CY volume, with the spread of target values shrinking with each generation as the GA hones in on the best DNA.

\begin{figure}[t!]
    \centering
    \includegraphics[width=0.9\linewidth]{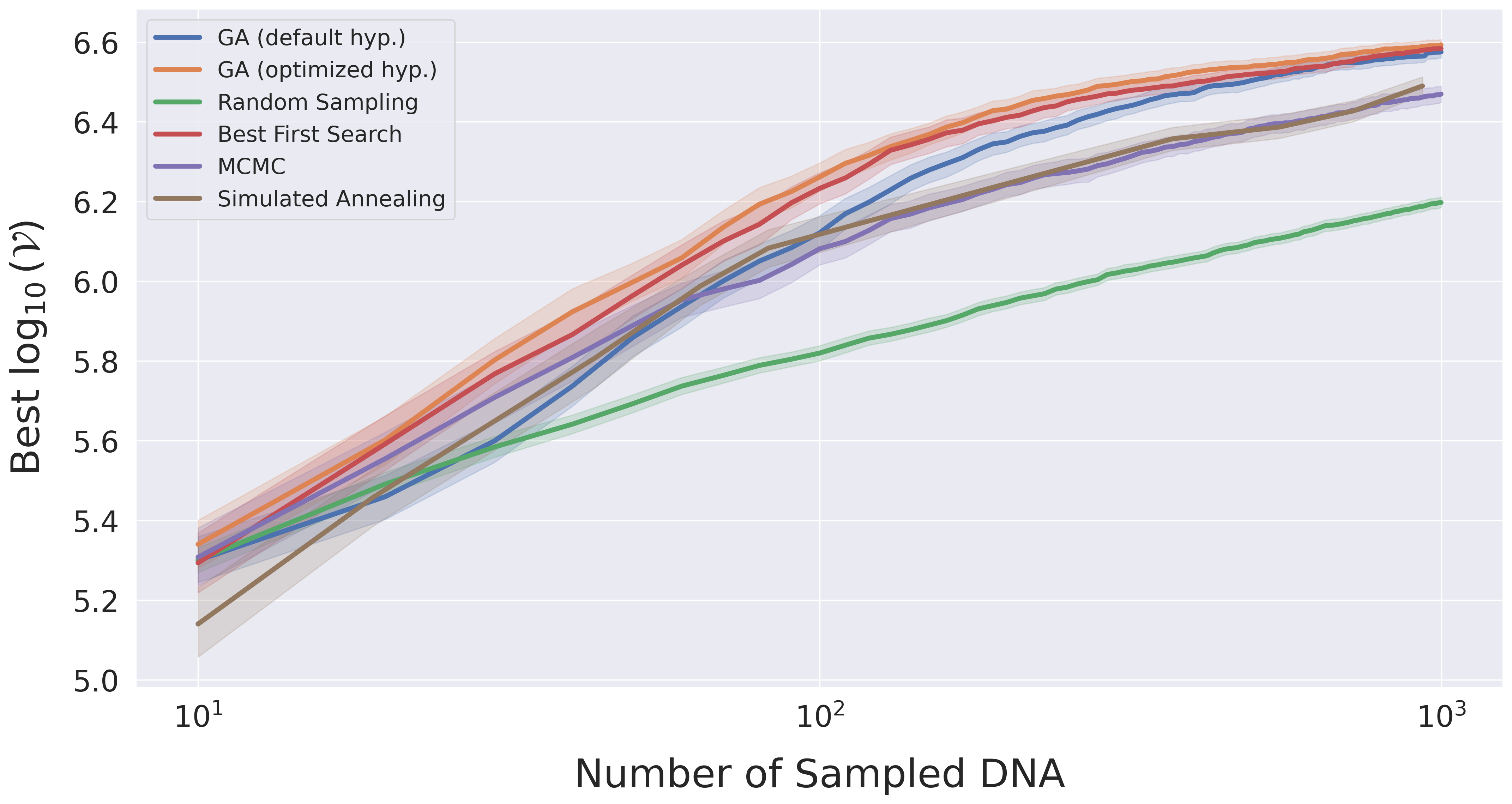}
    \caption{Performance comparison between different sampling and optimization algorithms when maximizing the Calabi-Yau volume. In particular, we plot the average best encountered target $\log_{10}(\mathcal{V})$ value as a function of unique encountered DNA. We see the DNA encoding of CYs enables many canonical optimization methods to noticeably outperform brute force search, with the GA and Best-First Search performing best. Moreover, we witness the non-trivial effect of Bayesian Optimization of hyperparameters.
    }
    \label{fig:models-h11-23}
\end{figure}

Next, let us quantify the effectiveness of the GA and the hyperparameter tuning described in \S\ref{sec:hyperopt}. In Fig.~\ref{fig:models-h11-23}, we compare the optimized GA with the default GA with generic hyperparameters (e.g., uniform preference between selection/crossover/mutation units)\footnote{We implement simulated annealing using the Python package \texttt{simanneal} \cite{simanneal} which only readily allows the user to specify the number of DNA encountered in a run, rather than the number of unique DNA. It is this limitation which leads to the curve endpoints for the simulated annealing in Fig.~\ref{fig:models-h11-23} not lining up perfectly with the remaining models' curves, where unique encountered DNA specification was readily accessible.}. In addition, we also apply other random sampling and optimization algorithms, namely:
\begin{itemize}
    \item \textbf{Random Sample} is a brute force method to randomly sample DNA $\cC$ specifying the triangulations of two-faces from a uniform distribution.
    \item \textbf{Markov Chain Monte Carlo} (MCMC) is a method used for sampling from complex probability distributions. It involves constructing a Markov chain that has the desired distribution as its equilibrium distribution. By iteratively transitioning between states of the chain according to a specified transition probability, MCMC generates samples that approximate the desired distribution. The samples produced by MCMC can be used for estimating expectations and probabilities in statistical inference and Bayesian analysis.
    \item \textbf{Simulated Annealing} (SA) is a probabilistic optimization technique inspired by the annealing process in metallurgy employed to discover global optima within expansive solution spaces. It starts with an initial solution and iteratively explores neighboring solutions, gradually decreasing the exploration range akin to cooling molten metal. By accepting worse solutions with a decreasing probability, it can escape local optima and converge towards a global optimum in the search space.
    \item \textbf{Best-First Search} (BeFS) is an ad hoc graph search algorithm modelled on depth-first search such that adjacent nodes are ordered randomly and greedily explored when their target value exceeds that of the current node. For graphs with few edges per node (such as this search space, where each node has only $4 \times 4 + 6 \times 4 = 40$ Hamming neighbors) this algorithm avoids local minima traps by simply retracing its steps backward and explores broadly by remembering where it has been, unlike MCMC and SA which have no memory and await a low-probability event to move to less preferred nodes.
\end{itemize}
As for the GA, individual moves in the other algorithms will be given by changing one of the two-face triangulation labels $c_i$. In this way, we are doing optimization on the graph whose nodes are DNA and whose edges connect DNA a Hamming distance of one away. We stress that the rough continuity we commented on earlier in this section entails that optimization methods beyond GAs and those enumerated here might prove effective on this graph of FRSTs.

In Fig.~\ref{fig:models-h11-23}, we plot the performance of the individual algorithms as a function of the number of sampled DNA which includes in particular those failing to extend to full FRSTs. Said differently, the $x$-axis counts the number of states visited by the various algorithms. On the $y$-axis, we plot the best feature found by the model, averaged over $150$ different runs.\footnote{We choose to plot the best value against unique evaluations of target function as a proxy for wall time instead of e.g. performance vs. generation. This is because these evaluations are the rate-limiting step, more so than, say, the selection/crossover/mutation process occurring each generation.} We observe that, rather unsurprisingly, the random sampling strategies shows the worst overall performance. The MCMC as well as SA show slightly better success in maximizing the volume. Interestingly, the maximization problem is most efficiently tackled by both the GA with or without optimized hyperparameters and Best-First Search.\footnote{We comment that while BeFS is a nice hyperparameter-less model for simple graphs such as this one, it is likely to scale poorly with $h^{1,1}$ and associated exponential increases in graph size/connectivity.}

\begin{figure}[t!]
    \centering
    \includegraphics[width=0.9\linewidth]{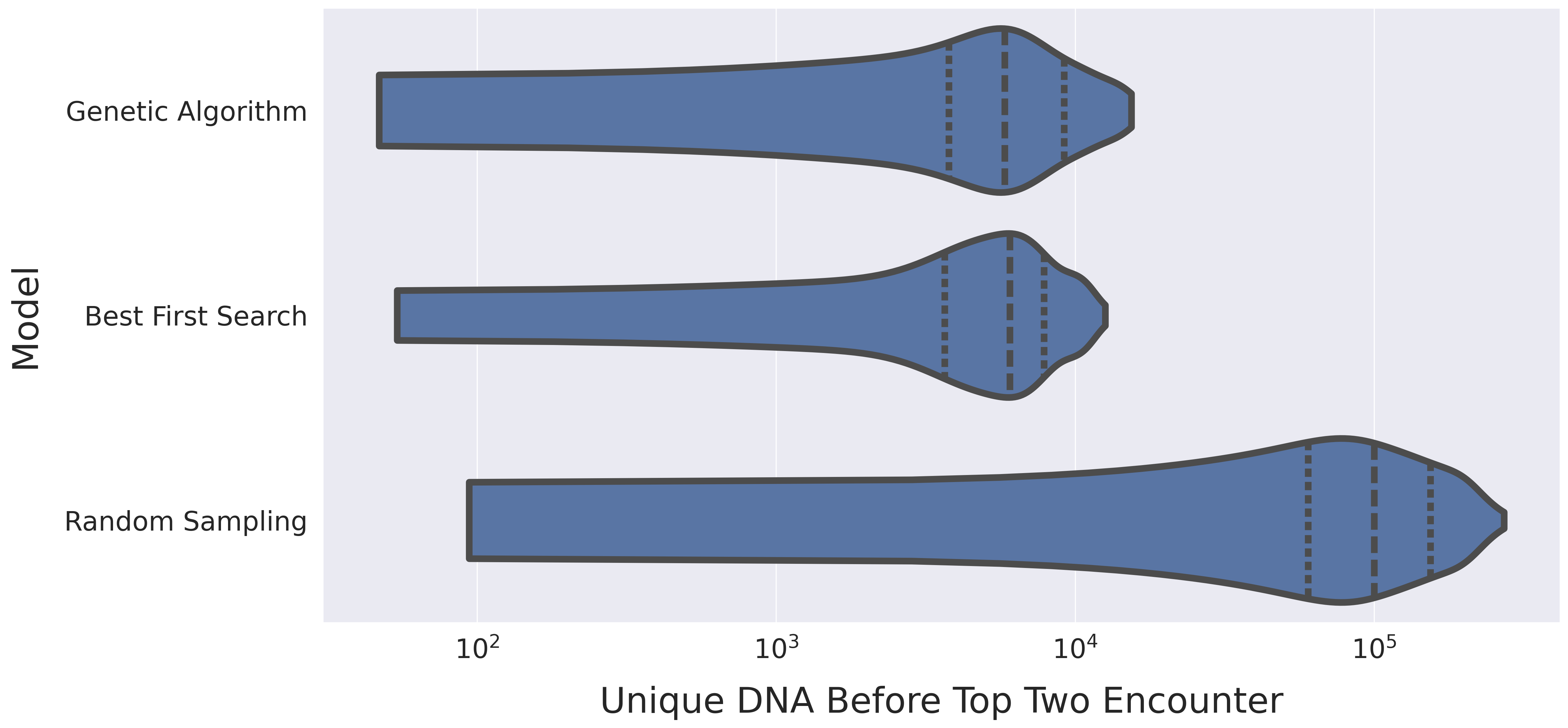}
    \caption{Comparison between number of unique DNA encountered before one of the two maximal values of $\log_{10}(\mathcal{V})$ is achieved for the two best optimization methods --- genetic algorithms and best-first search --- along with random sampling of DNA, for comparison purposes.}
    \label{fig:global_max}
\end{figure}

As discussed, a motivation for GAs are their ability to perform global optimization (i.e., not remain stuck within shallow local extrema). We exemplify this now, complementing our earlier discussion of the best DNA a model can find as a function of runtime (as measured by unique evaluations of $\log_{10}(\mathcal{V})$) with a brief analysis of the least runtime (i.e., fewest unique evaluated DNA) required to find the global maxima of our search space. Indeed, because our polytope permits enumeration of its DNA, we know the global structure of $\log_{10}(\mathcal{V})$ exactly. 

In particular, we identify that the two largest values of $\log_{10}(\mathcal{V})$ ($6.737$ and $6.713$) have associated DNA separated by a Hamming distance of $5$ (with the maximal Hamming distance being $8$ because the polytope has $8$ non-trivial two-faces). If we normalize the feature distance and Hamming distances by the ranges of their respective distributions, the feature distance is $\sim 100\times$ smaller, so we decide to treat these two DNA as degenerate global maxima and probe the number of unique encountered DNA $D$ required to find either one of them. In Fig.~\ref{fig:global_max} we present a violin plot comparing the distribution of $D$ for our two best models from the previous plot: the genetic algorithm and best-first search.\footnote{We stress that the GA used for Fig.~\ref{fig:models-h11-23} were designed to quickly locate the tail of the $\log_{10}(\mathcal{V})$ distribution but that their low genetic diversity (having a population size of $10$) and low mutation rate obstructs them from exploring the entire search space, as is necessary to find global extrema. To combat this, to produce Fig.~\ref{fig:global_max} we employ an ad hoc prescription where we increase the population size and mutation rate at two points in the training process to facilitate exploration. In particular, after 1000 unique evaluated DNA we increase the mutation rate and population size to $0.1$ and $100$, respectively, then after $4000$ unique evaluated DNA we increase to $0.2$ and $200$, respectively.} We see that the two methods perform comparably, with the GA having heavier tails on both ends than BeFS. Importantly, for both models one of the two best DNA are encountered with $\leq 16000$ unique DNA evaluated, meaning global extrema are reliably found with exposure to less than $5\%$ of the full search space.

To summarize, we established for the first time that GAs can tackle optimization problems defined on the space of NTFE FRSTs or, equivalently, on the space of homotopy inequivalent Calabi-Yau hypersurfaces. What is more, we provided evidence that the notion of DNA for such manifolds as introduced in \S\ref{sec:NTFEs} is a useful encoding not just for GAs, but more generally also for other optimization algorithms.\footnote{
For example, the utility of the two-face parameterization of triangulations in the context of Reinforcement Learning will be featured in the upcoming article \cite{ellirl}.
}

\subsection{Optimizing axion decay constants  --- $h^{1,1}=60$}\label{sec:example_h11_60}

As our second application, we study optimization problems for decay constants of axions \cite{Peccei:1977hh} which appear quite naturally in string compactifications \cite{Wen:1985jz,Svrcek:2006yi}. Typical EFTs obtained from Kaluza-Klein reductions of string theory contain hundreds of axions arising from the zero modes of higher-dimensional $p$-form potentials. Due to their prevalence, they have been argued to provide a rich testing ground for phenomenology in string theory \cite{Arvanitaki:2009fg} which is known as the \emph{string axiverse}.

In the present context, compactifications of Type IIB superstring theory on Calabi-Yau threefold hypersurfaces lead to the so-called \emph{Kreuzer-Skarke (KS) axiverse} \cite{Demirtas:2018akl}, see e.g. \cite{Mehta:2021pwf,Demirtas:2021gsq,Gendler:2023hwg,Gendler:2023kjt} for recent studies. We focus on axions $\phi^{a}$ descending from the Ramond-Ramond four-form $C_4$ whose action is of the form
\begin{equation}\label{eq:LagrangianAxion}
    \mathcal{L} = -\dfrac{M_{\text{pl}}^2}{2}\, K_{ab}(\partial_\mu\phi^{a})(\partial^\mu\phi^{b})-V(\phi)
\end{equation}
where $K_{ab}$ is the Kähler metric determined by classical geometric data. The leading-order scalar potential can be written as
\begin{equation}
    V(\phi) = \sum_I\, \Lambda_I^4 \biggl ( 1-\cos\left (2\pi\mathcal{Q}_{Ia}\phi^a\right )\biggl )
\end{equation}
and is induced by Euclidean D3-branes wrapping holomorphic four-cycles in the Calabi-Yau threefold. Here, we introduced the instanton charge matrix $\mathcal{Q}_{Ia}$, and the instanton mass scales $\Lambda_I$.

The mass spectrum of axions obtained from \eqref{eq:LagrangianAxion} typically spans several orders of magnitude due to hierarchies among the $\Lambda_I$ \cite{Arvanitaki:2009fg,Demirtas:2018akl,Mehta:2021pwf}. For convenience, we focus on the \emph{lightest} axion whose decay constant can be computed as
\begin{equation}
    f = M_{\text{pl}} \dfrac{\sqrt{\omega^a K_{ab}\omega^b}}{\sqrt{2}\pi\,  \mathcal{Q}_{b}\omega^b}\, .
\end{equation}
Here, $\omega^a$ is the generator of the complement orthogonal to the span of charge vectors for the $h^{1,1}-1$ smallest divisors (inducing the potential contributions for the heavier axions). Further, $\mathcal{Q}_{b}$ is the charge vector of the divisor setting the potential for the lightest axion. The value of $f$ depends again on the position in moduli space parametrized by the Kähler parameters $t^a$. As before, we evaluate the decay constants at the tip $t_{\star}^a$ of the stretched Kähler cone associated with a given FRST $\cT$.

In model building in string cosmology, one would like the value of $f$ to take values in a particular range. For instance, fuzzy dark matter usually requires $f\approx 10^{16}$ GeV (see e.g \cite{Cicoli:2021gss}) plus additional constraints on the mass of the axion. In the optimization language, this corresponds to solving the inverse problem: for a given target value $f_*$, which NTFE FRSTs $\cT$ give rise to decay constants $f(\cT,t_{\star}^a) \approx f_{*}$? Below, we demonstrate that such problems can be addressed with our GA. For our search, we use $f_* = 10^{14}$GeV as the target.

\begin{figure}[t!]
    \centering
    \includegraphics[width=0.7\columnwidth]{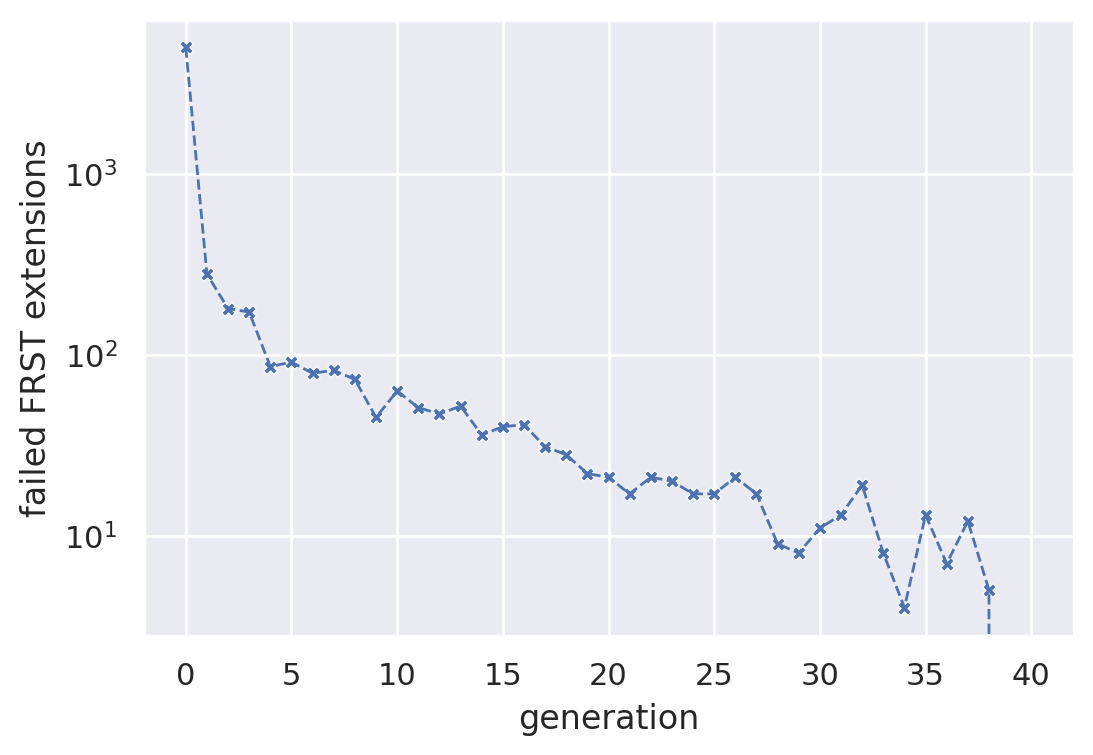}
    \caption{Number of failed extensions of two-face FRTs to FRSTs as a function of the generation.}
    \label{fig:h1160_failedlifts}
\end{figure}

To begin with, we choose a polytope $\Delta^\circ$ at $(h^{1,1},h^{1,2})=(60,4)$ whose vertices are given by
\begin{equation}
\left (
  \begin{array}{*{13}r}
  1 & 1 & 1 & 1 & -1 & -1 & 2 & -1 & -1 & -1 & -1 & -1 & -1\\
  0 & 2 & 0 & 0 & 0 & 0 & 1 & 2 & -2 & -2 & 0 & -2 & -2\\
  0 & 0 & 2 & 0 & 0 & -2 & 1 & 0 & -2 & -2 & 2 & 2 & 0\\
  0 & 0 & 0 & 2 & -2 & 0 & 1 & -2 & 0 & 2 & -2 & 0 & 2
\end{array} \right )\, .
\end{equation}
This polytope has 23 two-faces out of which 3, 6, 5, 3 and 6 have 1, 4, 64, 168, and 734 FRTs, respectively. Thus, the GA is again capable of exploring the entire space of NTFE FRSTs which is bounded by
\begin{equation}
    \#\text{NTFE FRSTs} \leq 4^6\times 64^5\times 168^3 \times 734^6\approx 3.3\times 10^{36}\, .
\end{equation}
Obviously, this is far too large to allow for a systematic enumeration. Again, we stress that the actual number of NTFE FRSTs is smaller since, as we demonstrate below, some configurations of two-face triangulations fail to extend to an FRST of $\Delta^\circ$, see Fig.~\ref{fig:h1160_failedlifts}.

\begin{figure}[t!]
    \centering
    \includegraphics[width=\columnwidth]{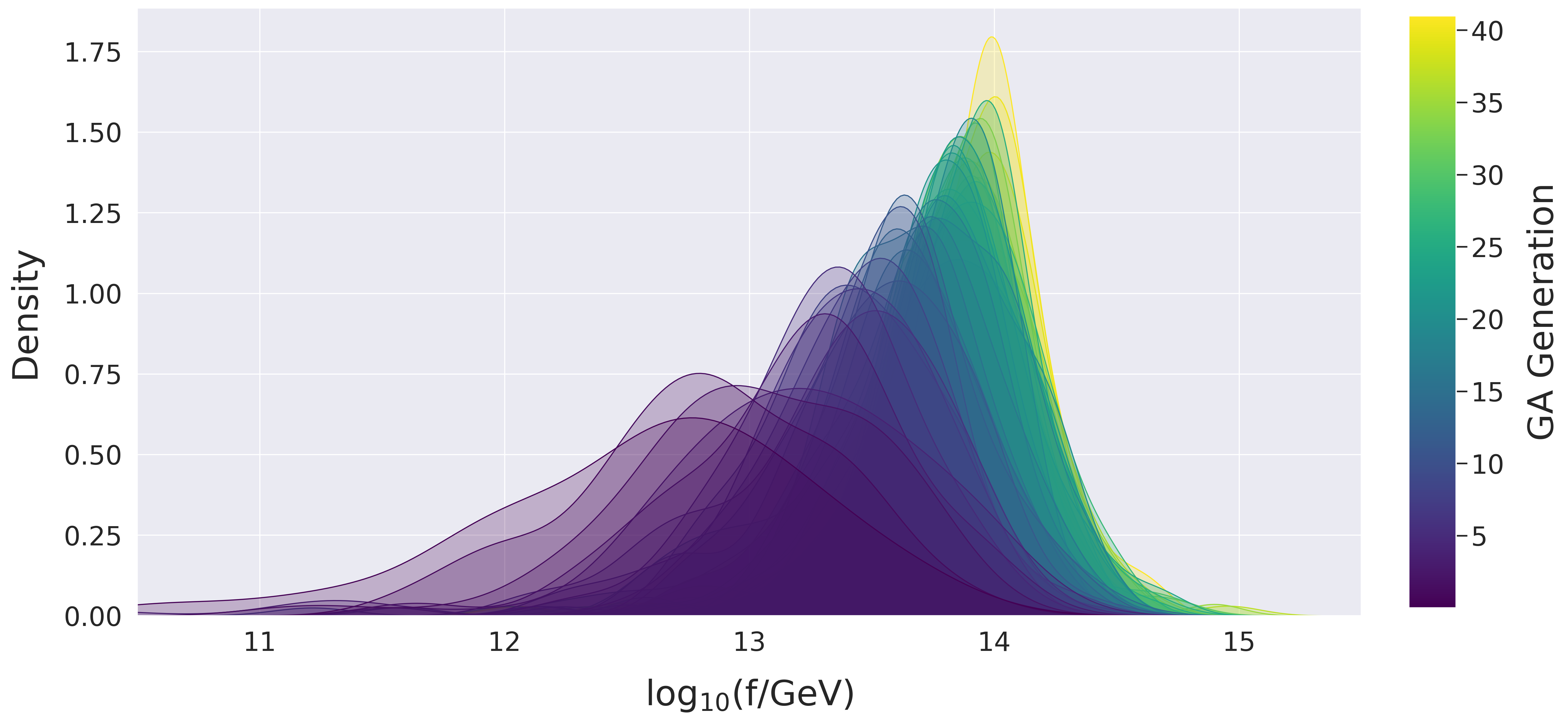}
    \caption{Distributions of decay constants for the lightest axion in each Calabi-Yau geometry colored by the generation.
    }
    \label{fig:h1160_results}
\end{figure}

We run the GA for $G=40$ generations with a constant population size of $P=100$.
For the fitness, we take a simple Gaussian
\begin{equation}
    F(f,f_*,\sigma) = \exp\left (-\dfrac{\log_{10}(f/f_{*})^2}{\sigma} \right )\, .
\end{equation}
Fig.~\ref{fig:h1160_results} shows the distributions of decay constants for the lightest axion in the corresponding Calabi-Yau geometry for each generation. The distributions quickly converge towards $f_*$ within the first few generations, though only after 30 generations the full distribution peaks around $f_*$. This trend is certainly expected since the GA initially learns general traits of the fittest FRSTs, while the convergence towards a global maximum of the fitness is only achieved at later stages.

\begin{figure}[t!]
    \centering
    \includegraphics[scale=0.395]{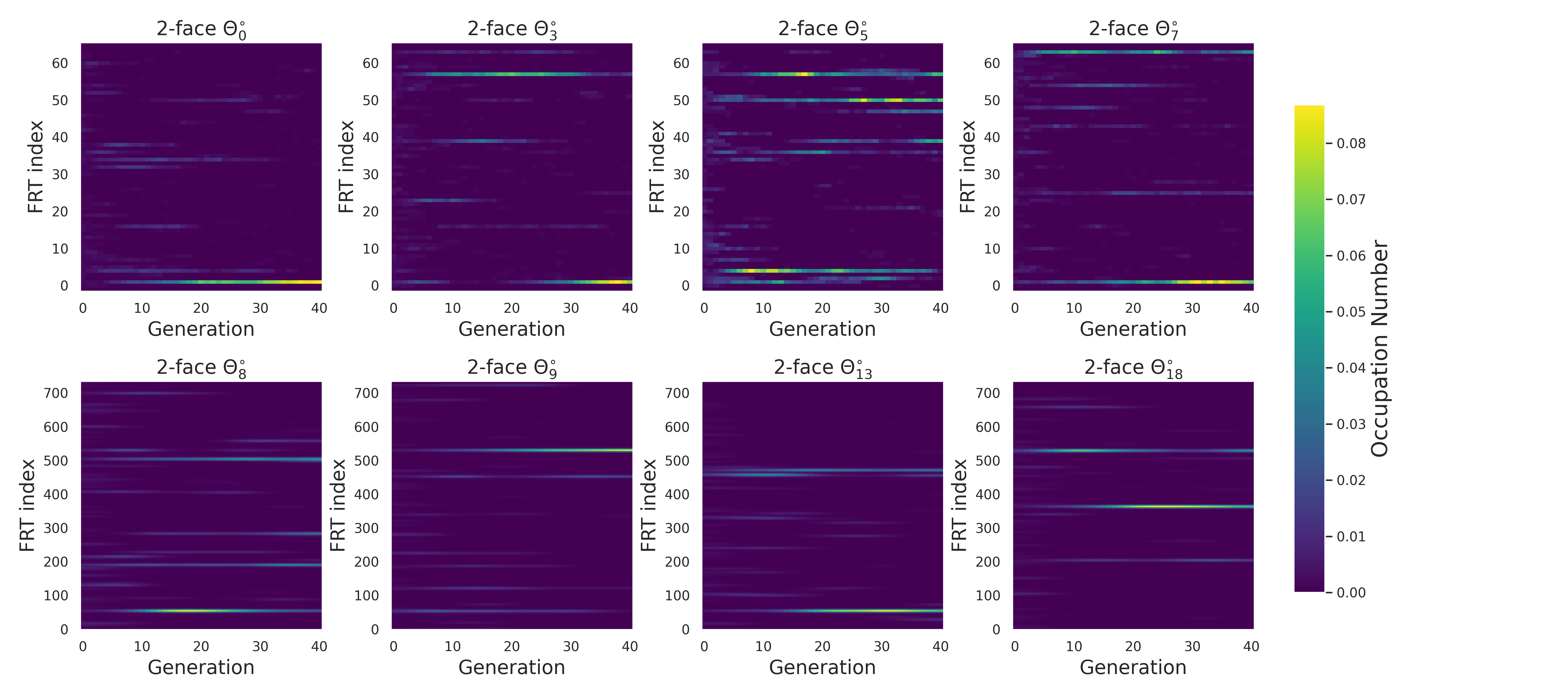}
    \caption{Occupation numbers of the two-face FRTs of eight selected two-faces where in the top row $\Theta^{\circ}_{0},\Theta^{\circ}_{3},\Theta_{5}^{\circ},\Theta^{\circ}_{7}$ have each 64 FRTs and in the bottom row $\Theta^{\circ}_{8},\Theta^{\circ}_{9},\Theta^{\circ}_{13},\Theta^{\circ}_{18}$ have 734 FRTs.}
    \label{fig:occ_2faces}
\end{figure}

To make this point clear, it is helpful to study the emergence of structures in the space of two-face triangulations associated with the specific task at hand similar to e.g. \cite{Cole:2019enn,Krippendorf:2021uxu,Cole:2021nnt} in the context of flux vacua. In Fig.~\ref{fig:occ_2faces}, we show the occupation number of the two-face FRTs of eight selected two-faces. While the initial population consisted of a random sample of FRTs, we observe interesting patterns during the GA's evolution. For instance, for the two-face $\Theta^{\circ}_{0}$, there is a single FRT that dominates the entire distribution after only around ten generations. This behaviour, where certain alleles are frozen at early times, is occasionally formalized in the context of \emph{schemata} \cite{Holland1975}, see also \cite{Abel:2014xta} for a detailed discussion.\footnote{Let us note that there is no clear consensus on the role of schemata in the GA literature, see in particular \S3.3 of \cite{reeves2002genetic} for a discussion.} The basic idea is that the GA detects certain genetic sequences as being beneficial in its strategy to maximize the fitness. This leads to certain parts of the chromosomes being predominantly fixed to a single value after only a handful of generations. In contrast, the spectrum of FRTs of $\Theta_{5}^\circ$ shows considerably more structure. Over the course of the GA's evolution, various different FRTs were populated and even dominated at intermediate stages, before the GA eventually settled at rather late times on roughly four triangulations dominating the final distribution.

Lastly, let us highlight the GA's effectiveness in finding novel NTFE FRSTs compared to random sampling of two-face FRTs. In Fig.~\ref{fig:h1160_failedlifts}, we plot the number of ``failed'' extensions of configurations of two-face FRTs as a function of the generation. Let us reiterate that extensions ``fail'' whenever the FRTs of two-faces cannot be consistently extended to a regular triangulation of $\Delta^\circ$. At generation $0$, we randomly sample FRTs of two-faces and need $4999$ choices $\cC$ of two-face triangulations to obtain the desired $P=100$ NTFE FRSTs. As becomes evident from Fig.~\ref{fig:h1160_failedlifts}, the number of such unextandable choices of two-face triangulations decreases rapidly across the entire duration of the search. Said differently, the GA develops a strategy to find new configurations of two-face triangulations that give rise to novel FRSTs. In fact, over the entire space of states visited by the GA, only $8.4\%$ of FRSTs are redundant, i.e., appear more than once. 

\subsection{Maximizing axion-photon couplings --- $h^{1,1}=491$}\label{sec:491}

As a final application of our GA, we study the largest polytope in the KS database with $(h^{1,1},h^{1,2})=(491,11)$, whose associated Calabi-Yau geometries are expected to dominate the entire KS set. Indeed, according to \cite{Demirtas:2020dbm}, the search space of NTFE FRSTs of this polytope is bounded by\footnote{The original bound of $\#\text{CYs}\lesssim 1.65 \times 10^{428}$ obtained in \cite{Demirtas:2020dbm} by estimating the number of two-face FRTs can be further improved to $\#\text{CYs}< 10^{296.1}$ \cite{macfaddenstepniczka} by exactly counting and using more stringent bounds on the number of two-face FRTs of the polytope with $h^{1,1}=491$.} $\# \text{CYs}\lesssim 1.65\times 10^{428}$. Because this polytope has both the largest search space in KS and features the most complex and computationally intensive geometries in this database, it represents an important and exciting target for optimization. Thanks to recent advances in computational geometry \cite{Demirtas:2022hqf} rendering quantities such as triple intersection numbers tractable even at $h^{1,1} = 491$, targeting this polytope with optimization techniques is now within reach.

The number of points in each of the ten two-faces of $\Delta^{\circ}$ is
\begin{equation}
 \left (\ell(\Theta^{\circ}_{0}),\ldots,\ell(\Theta^{\circ}_{9})\right ) =  (3,7,7,13,13,17,17,129,172,344)\, .
\end{equation}
For the two-faces with $17,13,7$ and $3$ points, we find respectively $19594$, $204$, $5$ and $1$ FRTs. There are three two-faces with more than $100$ points for which it is currently infeasible to compute the full set of FRTs.
We explored two approaches to these two-faces: the simplest was merely to choose a Delaunay triangulation for each one, and allow the GA to explore the two-face triangulations for all other two-faces. We additionally investigated the use of random subsamples of the triangulations for these two-faces (using methods such as the `fast' algorithm of \cite{Demirtas:2020dbm}), but found that this decreased extendability substantially, so we decided to use the former approach and defer a more comprehensive study of the DNA at $h^{1,1} = 491$ for future work. In summary, then, the GA's search space had size
\begin{equation}
    \#\text{NTFE FRSTs accessible to GA} \leq 5^2\times 204^2\times 19594^2 \approx 3.99\times 10^{14}\, .
\end{equation}
We note that this implies in particular that this search space is presumably smaller than the one for the polytope in \S\ref{sec:example_h11_60}: by restricting to the Delaunay triangulations of the three largest two-faces, we are significantly limiting the choices of DNA.

For this polytope, we selected axion-photon couplings $g_{a\gamma\gamma}$ as our target, motivated by the result that these are most vulnerable to phenomenological constraints at large $h^{1,1}$ \cite{Halverson:2019cmy,Gendler:2023kjt}.\footnote{By contrast, many other axionic observables are much further from constraints at large $h^{1,1}$: e.g., the misalignment production of dark matter and the QCD $\theta$ angle \cite{Demirtas:2021gsq,Gendler:2023kjt}. This is because many such observables scale with the decay constant $f$: see the ensuing discussion in the text for commentary on the loosely inverse relationship between $g_{a\gamma\gamma}$ and $f$.} Let us briefly review axion-photon couplings in string theory, following the notation and summarizing the findings of \cite{Gendler:2023kjt}, see e.g. \cite{Conlon:2007gk,Cicoli:2012sz,Halverson:2019kna,Halverson:2019cmy,Schallmoser:2021sba} for related work. The Lagrangian \eqref{eq:LagrangianAxion} is supplemented by
\begin{equation}
    \mathcal{L}_{a\gamma\gamma} = -\mathcal{Q}^{\text{EM}}_a\, \phi^a\, \dfrac{\alpha_{\text{EM}}}{4}\, F_{\mu\nu}\tilde{F}^{\mu\nu}\kom \tilde{F}^{\mu\nu} = \dfrac{1}{2}\epsilon^{\mu\nu\rho\sigma}F_{\rho\sigma}
\end{equation}
in terms of the fine structure constant $\alpha_{\text{EM}}$, the field strength of electromagnetism $F_{\mu\nu}$, and the charge $\mathcal{Q}^{\text{EM}}_a$ of the divisor hosting QED, see below. The axion-photon couplings $g_{a\gamma\gamma}$ naturally emerge upon a change of basis achieving a canonical kinetic term and lower-diagonal mass matrix and thus receive contributions from the charge matrix $\mathcal{Q}_{Ia}$, the mass scales $\Lambda_I$, and the K\"ahler metric $K_{ab}$. 

Conventionally, the decay constants $f$ are related to photon couplings by $g_{a\gamma\gamma} \leftrightarrow \frac{\alpha}{2\pi f}$, but this does not hold generically for the $C_4$ string axiverse. In particular, if one defines 
\begin{equation}
    C_\gamma = g_{a\gamma\gamma} \tfrac{2\pi f}{\alpha}
\end{equation}
to measure the deviation away from this pattern, one finds that $C_\gamma \ll 1$ and that it decreases with $h^{1,1}$, generically satisfying $C_\gamma \lesssim 10^{-20}$ at $h^{1,1} = 491$ \cite{Gendler:2023kjt}. This arises due to two compounding suppression effects, roughly given as follows. First, the typical hierarchies of instanton actions imply that photons have suppressed couplings to axions with larger instanton actions than that of the QED divisor (i.e., axions with mass below a \textit{light threshold}). Second, coupling suppression also results from the sparsity of off-diagonal elements of the Kähler metric (\textit{kinetic isolation}), a consequence of the infrequency of non-trivial prime toric divisor intersections (especially at large $h^{1,1}$). In conclusion, then, $g_{a\gamma\gamma}$ is a highly non-trivial function to optimize, even beyond the complexity of the axion decay constants studied in the previous section.

\begin{figure}[t!]
    \centering
    \includegraphics[width=0.95\linewidth]{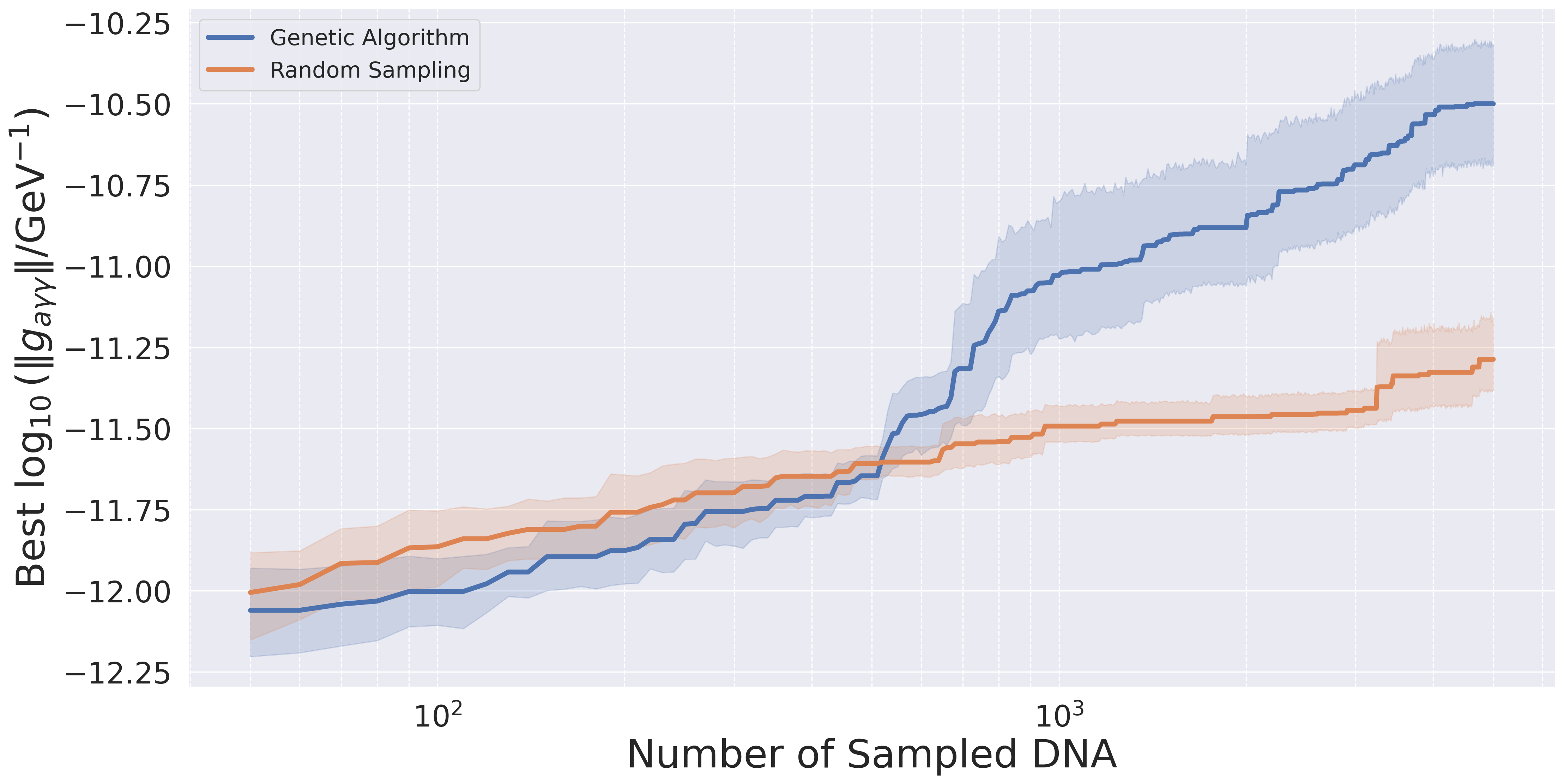}
    \caption{Best $\|g_{a\gamma\gamma}\|$ found as a function of number of unique DNA encountered while optimizing, averaged over $17$ GA runs and compared with average performance of random DNA sampling. The average maximum value of $\log_{10}(\|g_{a\gamma\gamma}\|/\text{GeV}^{-1})$ after $5000$ unique evaluated DNA is 
    $-10.5 \pm 0.22$ for the GA, and $-11.3 \pm 0.14$ for random sampling for $95\%$ confidence.
    }
    \label{fig:491}
\end{figure}

Our optimization task is to maximize an effective overall axion photon coupling, given by the Euclidean norm of the vector of axion-photon couplings for each of the $h^{1,1} = 491$ axions in our theory. We denote this $\|g_{a\gamma\gamma}\|$. For this to be well-defined, we must choose a divisor to host QED and a point in the K\"ahler cone: to pick these, we take some moderate inspiration from phenomenology. In particular, if we assume QCD and QED are hosted on separate divisors, the correct IR couplings are realized for divisor volumes of $40$ and $127.5$, respectively. We decide to compute the tip of the stretched Kahler cone and a basis of the smallest prime toric divisors, dilate until the smallest basis prime toric divisor $D$ has volume $40$ (thinking of this as our QCD divisor), and then place QED on the next smallest basis divisor $D'$ which intersects $D$, requiring that this divisor have volume less than $127.5$. This determines our point in moduli space and the QED divisor: if such a dilation cannot be performed, or such a QED candidate divisor does not exist, we set the target to be $-\infty$. In this way, our $g_{a\gamma\gamma}$ computation is mildly phenomenologically informed, but we stress that our optimized values should taken with a grain of salt: we omit many crucial checks of the legitimacy of the associated low-energy EFT.

We train GAs with target $\log_{10}\|g_{a\gamma\gamma}\|$ with population size $P = 50$ until each has evaluated $5000$ unique DNA. It takes $\mathcal{O}(\text{hours})$ on standard hardware for a single run, rendering our hyperparameter optimization methods --- for which we were performing hundreds of runs --- impractical. We compare the outcome of the GA with random sampling with error bars from averaging over $10$ runs in Fig.~\ref{fig:491}. The GA distinguishes itself from a brute force search approach after exposure to only $\sim 500$ unique DNA from the search space and continues to improve at a much faster rate than random sampling. In particular, within $5000$ unique seen DNA the GA has found effective axion-photon couplings over an order of magnitude larger than encountered by random sampling. To make this more quantitative, the average maximum value of $\|g_{a\gamma\gamma}\|$ found by the two methods after $5000$ unique evaluated DNA is 
\begin{equation}
    \log_{10}(\|g_{a\gamma\gamma}\|/\text{GeV}^{-1})_{\text{max}}=
    \begin{cases}
        -10.5 \pm 0.22 & \text{Genetic Algorithm} \, ,\\
        -11.3 \pm 0.14 & \text{Random Sampling}\, 
    \end{cases}
\end{equation}
for $95\%$ confidence. Thus, we find the encouraging result that, even without any kind of tuning or hyperparameter optimization, a GA can improve upon brute-force searching even for the polytope with the largest and most computationally intricate search space.

\section{Conclusions}\label{sec:conclusion}

In this paper, we demonstrated that Genetic Algorithms are a powerful tool to study the vast landscape of Calabi-Yau geometries in the KS database. Specifically, we implemented an algorithm acting on two-face triangulations for a fixed polytope to optimize objective functions defined on the space of Calabi-Yau threefold hypersurfaces. In fact, our encoding ensured that we were working only with FRSTs whose induced two-face triangulations are distinct. In this way, we removed a large fraction of the trivial redundancies arising when mapping polytope triangulations to Calabi-Yau geometries, recall Fig.~\ref{fig:ratioFRSTs}.

We presented three applications in the bulk of the paper. First, we studied a polytope at $h^{1,1}=23$ for which we were able to compute the full list of $331192$ NTFE FRSTs and the associated Calabi-Yau volumes at the tip of the stretched Kähler cone. We argued that the Calabi-Yau volume varies loosely continuously with the natural notions of distance --- Hamming and flip --- and features a funnel-like topography near maxima, suggesting the amenability of our DNA encoding to optimization methods like GAs. We demonstrated that our GA is outperforming other well-known sampling and optimization strategies like MCMC and Simulated Annealing. Crucially, we established that our encoding for Calabi-Yau threefold hypersurfaces in terms of DNA (the associated two-face triangulations) is extremely useful not just for GAs, but also for other optimization and sampling methods.

Afterwards, we studied a polytope at $h^{1,1}=60$ for which we computed all fine, regular two-face triangulations. Here, the search space of DNA for NTFE FRSTs was upper bounded by $\approx 3.3\times 10^{36}$ for which a systematic scan would be simply infeasible. For this polytope, we studied the distributions of decay constants $f$ associated with the lightest $C_4$-axion in compactifications of Type IIB string theory. We showed that the GA is efficient in locating FRSTs with axions having decay constants $f\approx f_*= 14$ GeV. This exercise serves as a proof of concept that GAs are indeed capable of solving inverse problems defined on the space of (NTFE) FRSTs. We highlighted that there are non-trivial structures arising in the space of two-face triangulations (see Fig.~\ref{fig:occ_2faces}) that deserve further scrutiny.

As a final application, we studied the polytope with $h^{1,1}=491$ which is believed to provide the majority of Calabi-Yau geometries in the KS database \cite{Demirtas:2020dbm} which are also computationally the most challenging. Because three of the two-faces are too large to enumerate all of their FRTs, we chose to fix a Delaunay triangulation for these two-faces and allow the GA to explore all triangulations for all other two-faces.
This effectively limited the search space accessible to the GA in our applications to $\lesssim 3.99\times 10^{14}$ states corresponding to NTFE FRSTs. For this polytope, we maximized axion-photon couplings at points in K\"ahler moduli space motivated by phenomenological considerations following \cite{Gendler:2023kjt}. Even in the absence of hyperparameter optimization --- which is obstructed by the steep $\|g_{a\gamma\gamma}\|$ evaluation time for this complex polytope --- we find that generic GAs lacking hyperparameter tuning significantly outperform random sampling after exposure to very few unique DNA, finding effective axion-photon couplings ten times as large, cf.~Fig.~\ref{fig:491}. Having found success in the maximal case, it follows that the entire KS database is within the reach of our optimization methods. We emphasize again however that we did not explore the space of two-face triangulations for the three largest two-faces of the polytope with $h^{1,1} = 491$: this is an important target for future work.

By implementing our methods in an open-source Python package \texttt{cyopt} \github{https://github.com/sheride/cyopt}, we hope to have provided a useful and accessible tool for approaching FRST sampling and optimization problems relevant for communities in string theory and mathematics.

There are multiple clear directions to extend the work done here and keep pushing the frontiers of optimization on toric Calabi-Yau hypersurfaces. First, one can apply methods studied in this paper toward more diverse and phenomenologically rich targets: for example, the engineering of stringy models for fuzzy dark matter --- necessitating the tuning of both masses and decay constants --- is potentially within reach (see e.g. \cite{Cicoli:2021gss}). 
Second, this paper considers only small subsets of the full theory space of Calabi-Yau string compactifications: we optimize over the Calabi-Yau geometries for fixed polytopes and standardized locations in moduli space. Thus, we would benefit greatly from the ability to additionally optimize over Kreuzer-Skarke polytopes and Calabi-Yau moduli spaces. In particular, the development of parametrizations for polytopes and for moduli spaces which are amenable to standard optimization methods (analogous to the DNA we have introduced here) would be extremely useful. Progress on the polytope front could be made by combining our methods with the tools of \cite{Berglund:2023ztk} for obtaining and optimizing reflexive polytopes with GAs. Improving the situation on the moduli space front is the topic of upcoming work by the authors \cite{dmss}. As progress is made in these directions, the objective driving string phenomenology --- discovering instances of phenomenological viability in the vast string landscape --- becomes increasingly tractable.

\subsubsection*{Acknowledgments}

We would like to thank Alex Cole, Mehmet Demirtas, Elli Heyes, Sven Krippendorf, Liam McAllister, Jakob Moritz, and Gary Shiu for discussions on related topics. AS is grateful to Alex Cole and Gary Shiu for earlier collaborations on related topics. We are grateful to Naomi Gendler for sharing code for computing axion-photon couplings. The research of NM, AS, and ES was supported in part by NSF grant PHY–2014071.

\appendix
\addtocontents{toc}{\protect\setcounter{tocdepth}{1}}

\bibliographystyle{utphys}
\bibliography{Literatur}

\end{document}